\documentclass[fleqn]{jaa}
\usepackage{natbib}

\usepackage[T1]{fontenc} 
\usepackage{ae,aecompl}
\usepackage{graphicx}	
\usepackage{amsmath}	
\usepackage{amssymb}	
\usepackage{textcomp}
\usepackage{verbatim}
\usepackage{longtable}
\usepackage{multirow}
\usepackage{caption}
\usepackage{subcaption}
\usepackage{multicol,lipsum}
\usepackage{mathtools, cuted}
\usepackage{bigints}
\usepackage[bottom]{footmisc}
\usepackage{hyperref}
\usepackage{float}
\usepackage{xcolor}

\usepackage{color}
\usepackage{soul}  
\usepackage{mwe}
\usepackage{array}
\usepackage{mathtools,euscript}

\begin{document}\sloppy
\title{Reconstruction of full sky CMB $\bf{E}$ and $\bf{B}$ modes spectra removing $\bf{E}$-to-$\bf{B}$ leakage from partial sky using deep learning}

\author{Srikanta Pal\textsuperscript{1} and Rajib Saha\textsuperscript{1,*}}
\affilOne{\textsuperscript{1}Department of Physics,\\ Indian Institute of Science Education and Research Bhopal,\\ Bhopal - 462066,\\ Madhya Pradesh, India}

\twocolumn[{

\maketitle

\corres{rajib@iiserb.ac.in}

\msinfo{\#\#\#\#}{\#\#\#\#}

\begin{abstract}
Incomplete sky analysis of cosmic microwave background (CMB) polarization spectra poses a major problem of leakage between $E$- and $B$-modes. We present a machine learning approach to remove this $E$-to-$B$ leakage using a convolutional neural network (CNN) in presence of detector noise. The CNN predicts the full sky $E$- and $B$-modes spectra for multipoles $2 \leq \ell \leq 384$ from the partial sky spectra for $N_{\rm{side}} = 256$. We use tensor-to-scalar ratio $r=0.001$ to simulate the CMB polarization maps. We train our CNN using $10^5$ full sky target spectra and an equal number of noise contaminated partial sky spectra obtained from the simulated maps. The CNN works well for two masks covering the sky area of $\sim 80\%$ and $\sim 10\%$ respectively after training separately for each mask. For the assumed theoretical $E$- and $B$-modes spectra, predicted full sky $E$- and $B$-modes spectra agree well with the corresponding target spectra and their means agree with theoretical spectra. The CNN preserves the cosmic variances at each multipole, effectively removes correlations of the partial sky $E$- and $B$-modes spectra, and retains the entire statistical properties of the targets avoiding the problem of so-called $E$-to-$B$ leakage for the chosen theoretical model.
\end{abstract}


}]


\doinum{\#\#\#\#}
\artcitid{\#\#\#\#}
\volnum{000}
\year{0000}
\pgrange{1--}
\setcounter{page}{1}
\lp{24}

\section{Inroduction}
\label{introduction}
The temperature and polarization anisotropies of the cosmic microwave background (CMB) radiation are the useful sources to extract the information of the early universe as well as the information of the present universe. This polarized CMB radiation was generated at the last scattering surface (LSS) with a temperature of $2970 \ \rm{K}$ approximately. In the present universe, we observe this CMB as a blackbody radiation of temperature $2.725\pm0.002 \ \rm{K}$~\citep{Fixsen_1996,Mather_1999}. It was first detected by~\cite{Penzias_1965}. The COBE satellite~\citep{Fixsen_1994,Bennett_1996} first measured the tiny temperature anisotropy of the CMB radiation. Then, the WMAP satellite~\citep{Hinshaw_2013,Bennett_2013} estimate the CMB angular power spectra of temperature and polarization anisotropies improving the accuracy of the previous measurements. Recently, the satellite-based experiment like Planck~\citep{PlanckI_2020,PlanckV_2020}, and the ground-based experiments like ACT~\citep{Sievers_2013,Choi_2020,Aiola_2020}, SPT~\citep{Hou_2014,Benson_2014,Dutcher_2021} released their more accurate analysis of the CMB observations. The major goal of the CMB community in the modern era is to measure the anisotropies of the CMB polarization more precisely, which can provide us the necessary information to understand the early universe accurately.

The CMB polarization can be explained by two useful polarization modes, i.e., $E$ and $B$-modes~\citep{Kamionkowski_1997,Zaldarriaga_1997,Hu_1997}. The density perturbations (scalar modes) in the early universe produce the $E$-mode polarization of even parity. There are two types of $B$-mode polarization of odd parity, i.e., primary and secondary $B$-modes. The primary $B$-mode can be generated by the primordial gravitational waves (tensor modes) during the cosmic inflation~\citep{Guth_1982} and the secondary $B$-mode is produced from the gravitational lensing of $E$-mode. Though the amplitude of the $B$-mode is much lower than the amplitude of the $E$-mode, the accurate detection of the $B$-mode polarization can explore the inflationary era of the early universe~\citep{Kamionkowski_2016}. The CMB polarization anisotropies was first detected by DASI~\citep{Leitch_2002,Kovac_2002}. Then, numerous experiments (e.g., WMAP~\citep{Hinshaw_2013,Bennett_2013}, QUaD~\citep{Brown_2009}, BICEP~\citep{Chiang_2010}, Planck~\citep{PlanckI_2020,PlanckV_2020}) measure specifically the $E$-mode polarization with high precision and constrain the early universe. For the better understanding of the early universe, the future experiments, e.g., Echo (aka CMB-Bharat\footnote{\url{http://cmb-bharat.in/}}), PIXIE~\citep{Kogut_2016}, CORE~\citep{Delabrouille_2018}, CCAT-prime\citep{Stacey_2018}, PICO\citep{Hanany_2019}, Lite-Bird\citep{Hazumi_2020}, attempt to observe the CMB polarization with higher sensitive instruments.

Observational CMB signal is contaminated by the different types of forground signal emitted from the local universe and the extragalactic point sources. To eliminate the effect of these forgrounds in the CMB, physicists study the partial sky CMB maps which are generated by applying the suitable binary mask in the full sky CMB maps. However, the partial sky analysis of the CMB polarization shows a major problem in the estimation of the partial sky $E$- and $B$-modes power spectra. The problem is that, at the time of the estimation of these polarization spectra from the partial sky polarization maps, a significant power of the $E$-mode spectrum is transmitted to the $B$-mode spectrum due to the presence of the ambiguous modes satisfying the properties of each of these two polarization modes. This transmission of the power from the $E$-mode spectrum to the $B$-mode spectrum is conventionally called $E$-to-$B$ leakage. It has been reported in literature by~\cite{Dodelson_2021}, the models of inflation~\citep{Guth_1982} can be very well constrained by measuring the weak primordial $B$-mode in CMB polarization. However, the measurement of this weak primordial $B$-mode from the partial sky observations may remain illusive unless we device a accurate method to negate the contribution of leakage from the strong $E$-mode to very weak $B$-mode which might give rise to spurious signal if not accounted for correctly.

There are numerous approaches attempted to effectively overcome the foregoing discussed problem of the leakage between $E$- and $B$-modes in the partial sky analysis. Working in harmonic space,~\cite{Lewis_2001} develop a specific window function for the clean separation of the $E$- and $B$-modes from the incomplete polarization sky map. Following the same harmonic-based method,~\cite{Lewis_2003} presented a general construction to achieve the separation of $E$- and $B$-modes for a arbitrary sky patches.~\cite{Bunn_2003} presented a classical procedure to distinguish the pure $E$, pure $B$ and ambiguous modes finding the orthonormal bases for all of these three components. Pure pseudo-$C_{\ell}$ estimator~\citep{Smith_2007,Zhao_2010} is also a traditional approach to distinguish the pure $E$- and $B$-modes from the ambiguous modes which causes the so-called $E$-to-$B$ leakage.~\cite{Kim_2010} proposed that, in the pixel space, the leakage can be reduce setting the lower value of tensor-to-scalar ratio ($r$) as well as reducing the sky fraction at the same time. In the pixel domain, another approach is the producing of the template of the leakage and applying this template in the corrupted map to clean the $E$-to-$B$ leakage~\citep{Liu_2019, Liu2_2019}. In these literature (i.e.,~\cite{Bunn_2017,Ramanah_2018,Ramanah_2019}), the authors employed the Wiener filtering method to separate the $E$- and $B$-modes polarization. However, each of these existing methods has the limitations in the performance depending upon the complexities of the implementation.

In the recent era, the machine learning (ML) approaches are applied in the various fields of physics including cosmology~\citep{Olvera_2021}. Artificial neural network (ANN) has the ability to reduce the computational time effectively for the Bayesian inference in cosmology~\citep{Graff_2012,Moss_2020,Hortua_2020,Gomez1_2021}.~\cite{Mancini_2022} show that Bayesian process can speed up using multiple types of ANN for estimating cosmological parameters using CMB data. ANNs are also used for the non-parametric reconstructions of cosmological functions~\citep{Escamilla_2020,Wang_2020,Dialektopoulos_2021,Gomez2_2021}. Using ANN, different types of foreground separation from CMB maps is performed by~\cite{Baccigalupi_2000}.~\cite{Petroff_2020} develop a neural network for cleaning foregrounds from CMB temperature anisotropies.~\cite{Munchmeyer_2019} use a neural network architecture, which is named WeinerNet by them, to speed up the Weiner filtering of partial sky CMB maps.~\cite{Shallue_2023} utilize convolutional neural network (CNN) to reconstruct cosmological initial conditions from the late-time non-linearly evolved density field.~\cite{Floss_2023} reconstruct the linear dark matter density field using neural network and improve the constrains on primordial non-Gaussianity.~\cite{Chanda_2021} construct a simple ANN to predict full sky CMB temperature anisotropy spectrum from partial sky CMB map for $N_{\rm{side}}=16$. In our eariler work~\citep{Pal_2023}, we develop an ANN to reconstruct full sky CMB temperature anisotropy spectrum from partial sky CMB map for $N_{\rm{side}}=256$.~\cite{Khan_2023} use ANN to detect dipole modulation in foregrund cleaned CMB temperature anisotrpoy maps provided by WMAP and Planck.  

In this article, we propose a ML approach to overcome the problem of the so-called $E$-to-$B$ leakage in the partial sky analysis of the CMB polarization in presence of detector noise. We develop a simple CNN system to predict the full sky $E$- and $B$-modes power spectra (both at the same time) using the corrupted partial sky $E$- and $B$-modes spectra as input in the network. We do not use any type of smoothing in the mask map which is used to create the partial sky $Q$ and $U$ polarization maps. We analyse this $E$-to-$B$ leakage problem in two cases using two binary masks. In the first case, we use a binary mask which creates the partial sky maps with approximately $80\%$ of the sky area. In the last case, we simply use a binary mask (containing only $\sim 10\%$ sky area) which eliminates approximately $90\%$ of the sky area including the galactic plane from the full sky maps. Due to the discontinuity at the edges of the masked region of the partial sky maps, some `ambiguous' modes are induced in the partial sky spectra obtained from these partial sky maps~\citep{Zhao_2010}. Taking care of these possible types of `ambiguous' modes as well as the noise biases, our CNN system can learn the hidden pattern between the corrupted partial sky spectra and the pure full sky spectra. Remembering this hidden pattern, this CNN system can predict the pure full sky spectra from the unknown partial sky spectra of the $E$- and $B$-modes polarization for the particular theoretical spectra used in our analysis. At the time of prediction by our trained CNN system, these input partial sky spectra are unseen to this trained CNN since these partial sky spectra were not used at the time of training procedure. Effectively removing the leakage between $E$- and $B$-modes spectra, these predictions of our CNN system preserve the all statistical properties of the full sky spectra of these polarization modes in case of each of these two binary masks.

We arrange the paper in the following. In section~\ref{stokes_sec}, we present a brief description about the Stokes parameter corresponding to the CMB radiation. In section~\ref{part_to_full_sec}, we describe the theoretical formalism of the full sky $E$- and $B$-modes power spectra from the partial sky spectra containing detector noises of these polarization modes. In section~\ref{th_spectra_sec}, we give the details of the theoretical $E$- and $B$-modes power spectra which are used in our analysis. The estimation procedure of the full sky polarization maps is discussed in section~\ref{full_QU_map_sec}. In section~\ref{noise_sec}, we describe the instrumental noises used in our analysis. In section~\ref{part_QU_map_sec}, we discuss about the estimation procedure of partial sky polarization maps. In section~\ref{full_spec_sec} and section~\ref{part_spec_sec}, we describe the simulations of the full sky and partial sky spectra of these CMB polarization modes. In section~\ref{cnn_sec}, we discuss the detailing of our CNN analysis to predict the leakage-free as well as noise-bias-free full sky $E$- and $B$-modes spectra. In section~\ref{results}, we discuss the obtained results from our CNN analysis. In section~\ref{realization_sec}, we present the predicted realization spectra of these polarization modes. In section~\ref{mean_std_sec}, we show the agreement between the mean predicted and mean target full sky spectra. We also show the agreement of the standard deviation of the predicted full sky spectra with the standard deviation of the target full sky spectra in the same section. In section~\ref{sign_sec}, we present the significance ratios of the CNN predicted full sky spectra of these polarization modes. In section~\ref{corr_sec}, we show the correlation matrices of the predicted full sky spectra comparing with the correlation matrices of the input cut-sky spectra. In this same section, we also show the correlations between target and predicted spectra of test set. Finally, we present the discussions and conclusions of our CNN analysis in section~\ref{discussions}.
\section{Formalism}
\label{formalism}
\subsection{Stokes parameters of CMB radiation}
\label{stokes_sec}
Polarized radiation is conventionally explained in terms of four fundamental Stokes parameters denoted by \textit{I, Q, U} and \textit{V}. In these Stokes parameters, the foremost one (\textit{I}) represents the intensity of the radiation at a observable position. Stokes parameters \textit{Q} and \textit{U} define the linear polarizations of the radiation in the standard cartesian coordinates ($\hat{x}, \hat{y}$) and the cartesian coordinates rotated by $45^{\circ}$ respectively. The last one (\textit{V}) of these Stokes parameters describes the polarization in circular basis ($\hat{l}=[\hat{x}+i\hat{y}]/\sqrt{2},\ \hat{r}=[\hat{x}-i\hat{y}]/\sqrt{2}$). 

In CMB measurements, the anisotropy is calculated as the fluctuation in the intensity, which is expressed as $\delta T(\theta,\phi) = \left[I(\theta,\phi)-\left<I\right>\right]T_0/\left<I\right>$ at the observable position ($\theta,\phi$), where $T_0$ and $\left<I\right>$ are the average temperature and average intensity of CMB radiation over the sky. This CMB anisotropy is a spin-$0$ (scalar) field. In CMB polarization experiments, radiation shows the linear polarization in terms of the measurements of $Q$ and $U$ Stokes parameters, which are not scalar fields. However, CMB does not show any polarization in the circular basis, which indicates the fourth Stokes parameter $V=0$ in case of CMB polarization measurements. Thus, the significant Stokes parameters corresponding to the CMB measurements are $I,Q$ and $U$, induced by scalar and tensor perturbations.
\subsection{Partial sky to full sky CMB polarization}
\label{part_to_full_sec}
The combinations of $Q$ and $U$ Stokes parameters, as spin-($\pm 2$) fields, are expressed as
\begin{eqnarray}
P_{\pm} &=& Q \pm iU \ .\label{P_pm}
\end{eqnarray}
These spin-($\pm 2$) fields, in terms of spin-weighted spherical harmonics, can be written as
\begin{eqnarray}
P_{\pm}(\hat{n}) & = & \sum_{\ell m} a_{\pm 2,\ell m} Y_{\pm 2,\ell m}(\hat{n}) \ , \label{P_sh}
\end{eqnarray}
where $\hat{n}$ denotes the observational position vector, $Y_{\pm 2,\ell m}(\hat{n})$ are the spin-weighted spherical harmonic functions and $a_{\pm 2,\ell m}$ are the expansion coefficients corresponding to these functions. Harmonic modes ($a_{\pm 2,\ell m}$) have $2 \ell +1$ degrees of freedom for a particular multipole $\ell$, since the index $m$ has the range from $-\ell$ to $\ell$.

Using the spherical harmonic coefficients from equation~\ref{P_sh}, the scalar $E$-mode and pseudo-scalar $B$-mode~\citep{Zaldarriaga_1997} corresponding to full sky CMB polarization map are stated as
\begin{eqnarray}
a_{E,\ell m}&=&-\frac{1}{2}(a_{2,\ell m}+a_{-2,\ell m}) \nonumber \\
&=&-\frac{1}{2}\int \Bigl[P_{+}(\hat{n})Y^{*}_{2,\ell m}(\hat{n}) + P_{-}(\hat{n})Y^{*}_{-2,\ell m}(\hat{n})\Bigr]d\Omega \ , \nonumber \\ \label{E_mode}\\
a_{B,\ell m}&=&\frac{i}{2}(a_{2,\ell m}-a_{-2,\ell m}) \nonumber \\ 
&=&\frac{i}{2}\int \Bigl[P_{+}(\hat{n})Y^{*}_{2,\ell m}(\hat{n}) - P_{-}(\hat{n})Y^{*}_{-2,\ell m}(\hat{n})\Bigr]d\Omega \ . \nonumber \\ \label{B_mode}
\end{eqnarray}
Moreover, the full sky angular power spectra corresponding to $E$- and $B$-modes are given by
\begin{eqnarray}
C^{EE}_{\ell} & = & \frac{1}{2\ell +1}\sum_{m}a_{E,\ell m}a^{*}_{E,\ell m} \ , \label{E_power} \\
C^{BB}_{\ell} & = & \frac{1}{2\ell +1}\sum_{m}a_{B,\ell m}a^{*}_{B,\ell m} \ . \label{B_power}
\end{eqnarray}

In absence of the leakage between $E$- and $B$-modes, as in case for full sky polarization, the scalar field corresponding to full sky $E$-mode and the pseudo-scalar field corresponding to full sky $B$-mode can be written as
\begin{eqnarray}
E(\hat{n}) & = & \sum_{\ell m} a_{E,\ell m}Y_{\ell m}(\hat{n}) \ , \label{E_field}\\
B(\hat{n}) & = & \sum_{\ell m} a_{B,\ell m}Y_{\ell m}(\hat{n}) \ , \label{B_field}
\end{eqnarray}
where $Y_{\ell m}(\hat{n})$ are the spherical harmonic functions corresponding to spin-$0$ field.

In presence of instrumental noises in smoothed $Q$ and $U$ polarization maps, the noise contaminated (or impure) spin-($\pm 2$) fields are expressed as
\begin{eqnarray}
P_{\pm}^{\rm{sm, imp}} &=& \left(Q^{\rm{sm}}+N_{Q}\right) \pm i\left(U^{\rm{sm}}+N_{U}\right) \ ,\label{P_sm_imp}
\end{eqnarray}
where $Q^{\rm{sm}}$ and $U^{\rm{sm}}$ are the polarization maps smoothed by using a selected FWHM (Full Width at Half Maximum). In equation~\ref{P_sm_imp}, $N_{Q}$ and $N_{U}$ are the instrumental noise maps corrsponding to $Q$ and $U$ polarizations respectively. Moreover, in spin-weighted spherical harmonic space, these smoothed and impure spin-($\pm 2$) fields can be expanded as
\begin{eqnarray}
P_{\pm}^{\rm{sm, imp}}(\hat{n}) &=& \sum_{\ell m} \left(a_{\pm 2,\ell m}^{\rm{sm}}+n_{\pm 2,\ell m}\right) Y_{\pm 2,\ell m}(\hat{n}) \ , \label{P_sm_imp_sh}
\end{eqnarray}
where $a_{\pm 2,\ell m}^{\rm{sm}}$ and $n_{\pm 2,\ell m}$ are the expansion coefficients corresponding to spin-weighted spherical harmonic functions for smoothed polarization maps and instrumental noise maps respectively.

From equation~\ref{P_sm_imp_sh}, the $E$- and $B$-modes corresponding to smoothed and pure full sky polarization maps are expressed as
\begin{eqnarray}
a_{E,\ell m}^{\rm{sm}} &=& -\frac{1}{2}\left(a_{2,\ell m}^{\rm{sm}}+a_{-2,\ell m}^{\rm{sm}}\right) \ = \ \mathcal{B}_{E}(\ell)a_{E,\ell m} \ , \label{E_mode_sm} \\
a_{B,\ell m}^{\rm{sm}} &=& \frac{i}{2}\left(a_{2,\ell m}^{\rm{sm}}-a_{-2,\ell m}^{\rm{sm}}\right) \ = \ \mathcal{B}_{B}(\ell)a_{B,\ell m} \ , \label{B_mode_sm}
\end{eqnarray}
where $\mathcal{B}_{E}(\ell)$ and $\mathcal{B}_{B}(\ell)$ are the beam window functions corresponding to $E$- and $B$-modes of CMB polarization respectively. In presence of instrumental noise, the smoothed as well as impure $E$- and $B$-modes for full sky polarizations can be written as
\begin{eqnarray}
a_{E,\ell m}^{\rm{sm,imp}} &=& \mathcal{B}_{E}(\ell)a_{E,\ell m}+n_{E,\ell m} \ , \label{E_mode_sm_imp}\\
a_{B,\ell m}^{\rm{sm,imp}} &=& \mathcal{B}_{B}(\ell)a_{B,\ell m}+n_{B,\ell m} \ , \label{B_mode_sm_imp}
\end{eqnarray}
where $n_{E,\ell m}$ and $n_{B,\ell m}$ are $E$- and $B$-modes corresponding to instrumental noise maps. These $E$- and $B$-modes for instrumental noise are expressed as
\begin{eqnarray}
n_{E,\ell m}&=&-\frac{1}{2}(n_{2,\ell m}+n_{-2,\ell m}) \ , \label{E_mode_noise}\\
n_{B,\ell m}&=&\frac{i}{2}(n_{2,\ell m}-n_{-2,\ell m}) \ , \label{B_mode_noise}
\end{eqnarray}
Assuming the uncorrelation between the noise maps $N_{Q}$ and $N_{U}$ as well as using the orthogonality conditions of $Y_{\pm 2,\ell m}$, the statistical properties of noise~\citep{Zaldarriaga_1997} are given by
\begin{eqnarray}
\bigl<n_{2,\ell m}n_{2,\ell' m'}^{*}\bigr> &=& \frac{8\pi \sigma_{P}^{2}}{N_{\rm{pix}}}\delta_{\ell\ell'}\delta_{mm'} \ , \label{noise2_auto} \\
\bigl<n_{-2,\ell m}n_{-2,\ell' m'}^{*}\bigr> &=& \frac{8\pi \sigma_{P}^{2}}{N_{\rm{pix}}}\delta_{\ell\ell'}\delta_{mm'} \ , \label{noise-2_auto} \\
\bigl<n_{2,\ell m}n_{-2,\ell' m'}^{*}\bigr> &=& \bigl<n_{2,\ell m}^{*}n_{-2,\ell' m'}\bigr> \ = \ 0 \ , \label{noise_cross}
\end{eqnarray}
where $\sigma_{P}$ defines the rms noise in $Q$ and $U$ polarization maps, and $N_{\rm{pix}}$ represents the number of pixel. Using equations~\ref{noise2_auto},~\ref{noise-2_auto} and~\ref{noise_cross}, the statistical properties of $E$- and $B$-modes of instrumental noise can be written as
\begin{eqnarray}
\bigl<n_{E,\ell m}n_{E,\ell' m'}^{*}\bigr> &=& \bigl<N_{\ell}^{EE}\bigr>\delta_{\ell\ell'}\delta_{mm'} \label{noiseE_auto} \\
\bigl<n_{B,\ell m}n_{B,\ell' m'}^{*}\bigr> &=& \bigl<N_{\ell}^{BB}\bigr>\delta_{\ell\ell'}\delta_{mm'} \ , \label{noiseB_auto} \\
\bigl<n_{E,\ell m}n_{B,\ell' m'}^{*}\bigr> &=& \bigl<n_{E,\ell m}^{*}n_{B,\ell' m'}\bigr> \ = \ 0 \ , \label{noiseEB_cross}
\end{eqnarray}
where $\bigl<N_{\ell}^{EE}\bigr>=\bigl<N_{\ell}^{BB}\bigr>=4\pi \sigma_{P}^{2}/N_{\rm{pix}}=constant$.

After removing the smoothing effect from full sky $E$- and $B$-modes, equations~\ref{E_mode_sm_imp} and~\ref{B_mode_sm_imp} can be expressed as
\begin{eqnarray}
a_{E,\ell m}^{\rm{imp}} &=& a_{E,\ell m}+n'_{E,\ell m} \ , \label{E_mode_imp}\\
a_{B,\ell m}^{\rm{imp}} &=& a_{B,\ell m}+n'_{B,\ell m} \ , \label{B_mode_imp}
\end{eqnarray}
where $n'_{E,\ell m} = \frac{n_{E,\ell m}}{\mathcal{B}_{E}(\ell)}$ and $n'_{B,\ell m} = \frac{n_{B,\ell m}}{\mathcal{B}_{B}(\ell)}$. In equations~\ref{E_mode_imp} and~\ref{B_mode_imp}, $a_{E,\ell m}^{\rm{imp}}$ and $a_{B,\ell m}^{\rm{imp}}$ define the noise contaminated (or impure) full sky $E$- and $B$-modes respectively after removing the beam effect from smoothed full sky $E$- and $B$-modes. Therefore, the impure spin-($\pm 2$) fields, related to $a_{E,\ell m}^{\rm{imp}}$ and $a_{B,\ell m}^{\rm{imp}}$, can be written as
\begin{eqnarray}
P_{\pm}^{\rm{imp}}(\hat{n}) &=& \sum_{\ell m} \left(a_{\pm 2,\ell m}+n'_{\pm 2,\ell m}\right) Y_{\pm 2,\ell m}(\hat{n}) \nonumber \\
&=& \sum_{\ell m} a^{\rm{imp}}_{\pm 2,\ell m} Y_{\pm 2,\ell m}(\hat{n}) \ , \label{P_imp_sh}
\end{eqnarray}
where $a_{\pm 2,\ell m}$ is related to $a_{E,\ell m}$ and $a_{B,\ell m}$, and $n'_{\pm 2,\ell m}$ is related to $n'_{E,\ell m}$ and $n'_{B,\ell m}$.

In case of the partial sky CMB polarization analysis, the $E$ and $B$ harmonic modes corresponding to impure partial sky polarization maps (from equation~\ref{P_imp_sh}) can be expressed as
\begin{eqnarray}
\tilde{a}_{E,\ell m} & = & -\frac{1}{2}\int \Bigl[P^{\rm{imp}}_{+}(\hat{n})W(\hat{n})Y^{*}_{2,\ell m}(\hat{n}) + \nonumber \\ 
&& \hspace{30pt} P^{\rm{imp}}_{-}(\hat{n})W(\hat{n})Y^{*}_{-2,\ell m}(\hat{n})\Bigr]d\Omega \ , \label{E_mode_part}\\
\tilde{a}_{B,\ell m} & = & \frac{i}{2}\int \Bigl[P^{\rm{imp}}_{+}(\hat{n})W(\hat{n})Y^{*}_{2,\ell m}(\hat{n}) - \nonumber \\ 
&& \hspace{30pt} P^{\rm{imp}}_{-}(\hat{n})W(\hat{n})Y^{*}_{-2,\ell m}(\hat{n})\Bigr]d\Omega \ , \label{B_mode_part}
\end{eqnarray}
where $W(\hat{n})$ is the finite window function applying on full sky $Q$ and $U$ polarization maps to generate partial sky $Q$ and $U$ maps. The window function, in spin-weighted spherical harmonic space of spin-($\pm 2$) field, is given by
\begin{eqnarray}
W^{(\pm 2)}_{\ell m\ell' m'} & = & \int Y_{\pm 2,\ell' m'}(\hat{n})W(\hat{n})Y_{\pm 2,\ell m}^{*}(\hat{n})d\Omega \ . \label{window}
\end{eqnarray}
Applying equations~\ref{P_imp_sh} and~\ref{window} in equations~\ref{E_mode_part} and~\ref{B_mode_part}, the impure partial sky $E$ and $B$ harmonic modes, in terms of the impure full sky $E$ and $B$ harmonic modes, can be expressed as
\begin{eqnarray}
\tilde{a}_{E,\ell m}  &=&  \sum_{\ell' m'}\Bigl[K^{(+)}_{\ell m\ell' m'}a^{\rm{imp}}_{E,\ell' m'}+iK^{(-)}_{\ell m\ell' m'}a^{\rm{imp}}_{B,\ell' m'}\Bigr] \ , \label{E_mode_part_to_full} \\
\tilde{a}_{B,\ell m}  &=&  \sum_{\ell' m'}\Bigl[K^{(+)}_{\ell m\ell' m'}a^{\rm{imp}}_{B,\ell' m'}-iK^{(-)}_{\ell m\ell' m'}a^{\rm{imp}}_{E,\ell' m'}\Bigr] \ , \label{B_mode_part_to_full}
\end{eqnarray}
where $K^{(\pm)}_{\ell m\ell' m'}$ are known as the mixing kernels~\citep{Ferte_2013}. These kernels (\ref{append_kernel}) are defined as
\begin{eqnarray}
K^{(\pm)}_{\ell m\ell' m'} &=& \frac{1}{2}\Bigl[W^{(+2)}_{\ell m\ell' m'}\pm W^{(-2)}_{\ell m\ell' m'}\Bigr] \ . \label{mix_kernel}
\end{eqnarray}
We note that the partial sky $E$- and $B$-modes are coupled with the full sky $B$- and $E$-modes through the mixing kernel $K^{(-)}_{\ell m\ell' m'}$, which is the source of the leakage between $E$- and $B$-modes in case of partial sky CMB polarization.
\begin{figure*}[h!]
\centering
\includegraphics[scale=0.4]{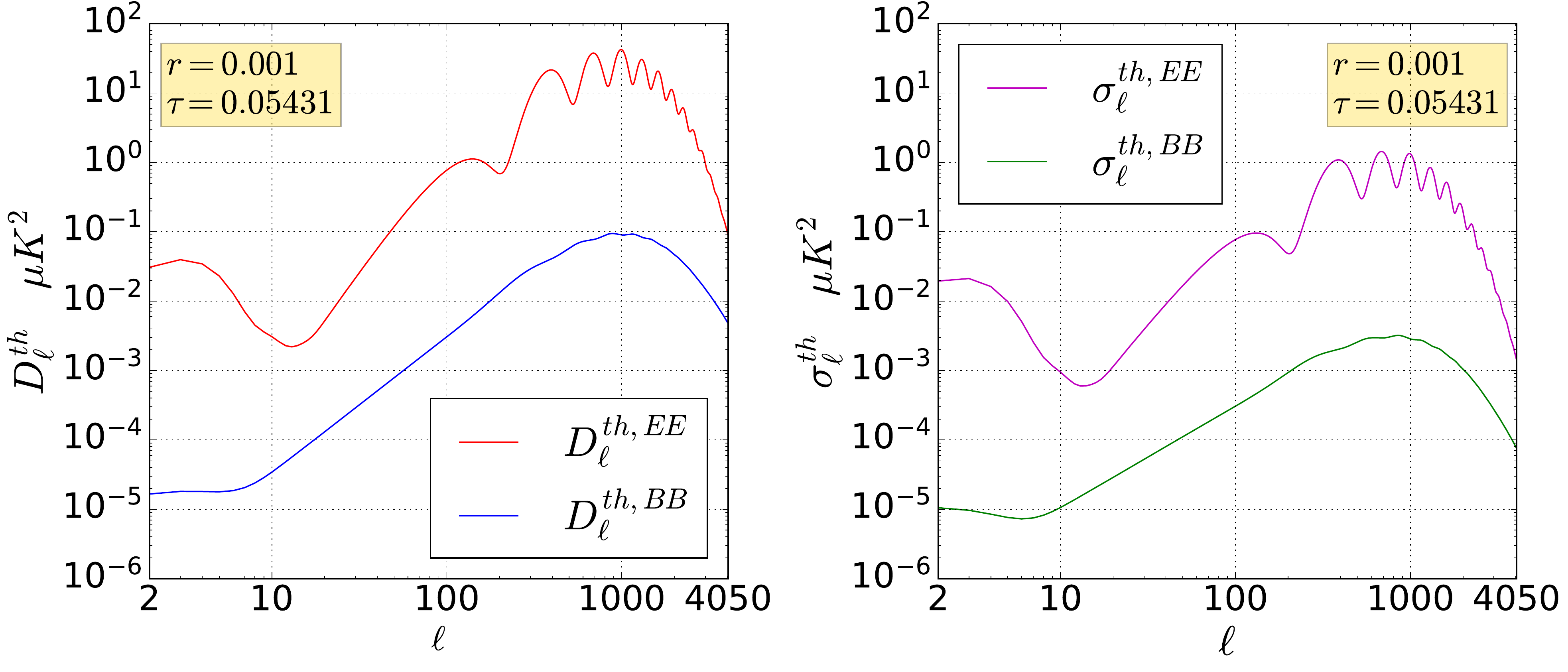}
\caption{Left panel of the figure shows the theoretical power spectra ($D_{\ell}^{th,EE}$ and $D_{\ell}^{th,BB}$) in $\mu K^2$, corresponding to the CMB $E$- and $B$-modes polarizations with maximum multipole $\ell_{\rm{max}}=4050$ and tensor-to-scalar ratio $r=0.001$, where $D_{\ell}^{th}=\ell(\ell +1)C_{\ell}^{th}/2\pi$. In the right panel, we show the cosmic standard deviations ($\sigma_{\ell}^{th}=\sqrt{2/(2\ell +1)} \ D_{\ell}^{th}$) in $\mu K^2$, corresponding to these spectra for the same values of $\ell_{\rm{max}}$ and $r$. Ignoring monopole ($\ell =0$) and dipole ($\ell =1$), both axes of each panel of the figure are represented in $\log_{10}$ scale.}
\label{theory_EB}
\end{figure*}

In presence of instrumental (or detector) noises, the ensemble averages of partial sky power spectra corresponding to $E$- and $B$-modes, in terms of the ensemble averages of full sky spectra of these modes, are given by
\begin{eqnarray}
\begin{pmatrix}
\bigl<\tilde{C}_{\ell}^{EE}\bigr> \\[5pt] \bigl<\tilde{C}_{\ell}^{BB}\bigr>
\end{pmatrix}
\ = \
\sum_{\ell'}
\begin{pmatrix}
M_{\ell \ell'}^{(+)} & M_{\ell \ell'}^{(-)}\\[5pt]
M_{\ell \ell'}^{(-)} & M_{\ell \ell'}^{(+)}
\end{pmatrix}
\begin{pmatrix}
\bigl<C_{\ell'}^{EE}\bigr>+\frac{\bigl<N_{\ell'}^{EE}\bigr>}{\mathcal{B}_{E}^{2}(\ell')} \\[5pt] \bigl<C_{\ell'}^{BB}\bigr>+\frac{\bigl<N_{\ell'}^{BB}\bigr>}{\mathcal{B}_{B}^{2}(\ell')}
\end{pmatrix} \ , \label{EB_power_part}
\end{eqnarray}
where $M_{\ell \ell'}^{(\pm)}$ denotes the mixing matrices corresponding to the CMB polarization~\citep{Alonso_2019}. These mixing matrices (\ref{append_matrix}) are defined by
\begin{eqnarray}
M_{\ell \ell'}^{(\pm)} &=& \sum_{mm'}\frac{1}{2\ell +1}|K^{(\pm)}_{\ell m\ell' m'}|^2 \ . \label{mix_matrix}
\end{eqnarray}
From the combined matrix representation (shown in equation~\ref{EB_power_part}) of the partial sky spectra of $E$- and $B$-modes, the full sky spectra of these polarization modes can be written as
\begin{eqnarray}
\begin{pmatrix}
\bigl<C_{\ell}^{EE}\bigr> \\[5pt] \bigl<C_{\ell}^{BB}\bigr>
\end{pmatrix}
\hspace{-5pt}&=& \hspace{-5pt}
\sum_{\ell'}
\begin{pmatrix}
M_{\ell \ell'}^{(+)} & M_{\ell \ell'}^{(-)}\\[5pt]
M_{\ell \ell'}^{(-)} & M_{\ell \ell'}^{(+)}
\end{pmatrix}^{-1}
\begin{pmatrix}
\bigl<\tilde{C}_{\ell'}^{EE}\bigr> \\[5pt] \bigl<\tilde{C}_{\ell'}^{BB}\bigr>
\end{pmatrix} \ - \
\begin{pmatrix}
\frac{\bigl<N_{\ell}^{EE}\bigr>}{\mathcal{B}_{E}^{2}(\ell)} \\[5pt]
\frac{\bigl<N_{\ell}^{BB}\bigr>}{\mathcal{B}_{B}^{2}(\ell)}
\end{pmatrix} \ . \nonumber \\ \label{EB_power_full}
\end{eqnarray}
This is the analytical expression of the relation between the partial sky (with detector noises) and full sky polarization power spectra at the level of ensemble averages. We configure this relation in a different way using a simple CNN system to learn the mapping from the partial sky power spectra (in presence of detector noise) corresponding to $E$- and $B$-modes to the full sky spectra of these CMB polarization modes.
\section{Methodology}
\label{methodology}
\subsection{Theoretical power spectra of $E$- and $B$-modes}
\label{th_spectra_sec}
\begin{figure*}[h!]
\centering
\includegraphics[scale=0.5]{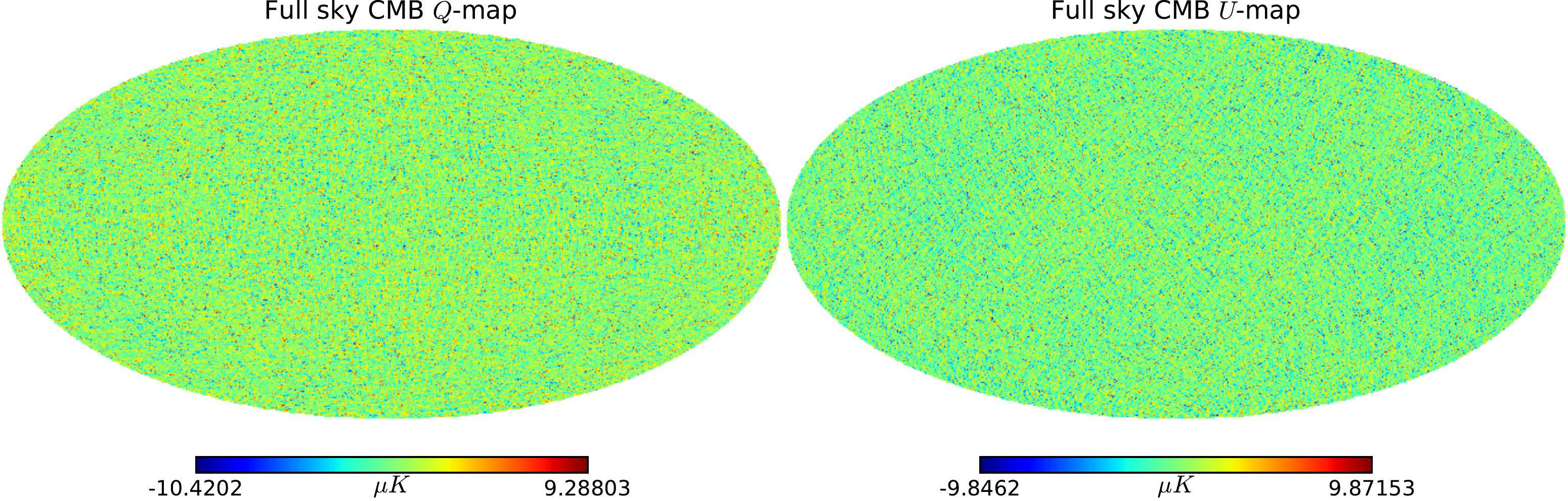}
\caption{Left panel of the figure shows the full sky $Q$ polarization map of CMB and the right panel of the figure represents the full sky $U$ polarization map of CMB. These noise free polarization maps are simulated for the resolution parameter $N_{\rm{side}}=256$ from the theoretical $E$- and $B$-modes spectra ($C_{\ell}^{th,EE}$ and $C_{\ell}^{th,BB}$) for a randomly selected seed value. Colour bar of each panel of this figure is in $\mu K$ unit.}
\label{QU_map_full}
\end{figure*}
We utilize the python-wrapper CAMB\footnote{\url{https://camb.readthedocs.io/en/latest/}} to generate the theoretical power spectra ($C_{\ell}^{th,EE}$ and $C_{\ell}^{th,BB}$) corresponding to CMB $E$- and $B$-modes. We use six cosmological parameters obtained by~\cite{PlanckVI_2020} and  a possible non-zero value of tensor-to-scalar ratio ($r$ ;~\cite{Tristram_2021}) to estimate these theoretical spectra. Let us define $D_{\ell} = \ell(\ell +1)C_{\ell}/2\pi$. In Figure~\ref{theory_EB}, we show these theoretical spectra ($D_{\ell}^{th,EE}$ and $D_{\ell}^{th,BB}$) in the left panel. In the right panel of this figure, we present the cosmic standard deviations ($\sigma_{\ell}^{th}=\sqrt{2/(2\ell +1)} \ D_{\ell}^{th}$) corresponding to these spectra obtained by CAMB for the maximum multipole $\ell_{\rm{max}}=4050$ and $r=0.001$. We ignore the monopole ($\ell =0$) and dipole ($\ell =1$) terms corresponding to these spectra. We note that the theoretical $B$-mode spectrum contains all possible modes, such as scalar,tensor and lensing. Moreover, in Table~\ref{cosmo_param}, we present the cosmological parameters, which are used for the estimation of these theoretical spectra.
\begin{table}[h!]
\begin{center}
\caption{Cosmological parameters~\protect\citep{PlanckVI_2020}, used for the estimation of the theoretical power spectra corresponding to CMB $E$- and $B$-modes, are presented in this table.}
\label{cosmo_param}
\begin{tabular}{ll}
\hline\hline
Parameter & Value \\ 
\hline
$\Omega_{b}h^{2}$ & $0.02238$ \\
$\Omega_{c}h^{2}$ & $0.1201$ \\
$H_{0}$ & $67.32$ \\
$\tau$ & $0.05431$ \\
$n_{s}$ & $0.966$ \\
$A_{s}$ & $2.1 \times 10^{-9}$ \\
$r$ & $0.001$ \\ 
\hline
\end{tabular}
\end{center}
\tablenotes{$\Omega_{b}h^2$ denotes today's baryonic density parameter, $\Omega_{c}h^2$ is today's density parameter of cold dark matter, $H_{0}$ is today's Hubble parameter in units of $\rm{km/sec/Mpc}$, $\tau$ represents the optical depth to decoupling surface, $n_{s}$ is the scalar spectral index, $A_{s}$ is the characterize parameter for the amplitude of initial perturbations and $r$ denotes the tensor-to-scalar ratio for CMB polarization.}
\end{table}
\subsection{Full sky $Q$ and $U$ maps of CMB polarization}
\label{full_QU_map_sec}
We use the python version (\texttt{healpy}\footnote{\url{https://github.com/healpy/healpy}}) of HEALPix\footnote{\url{https://healpix.sourceforge.io/}} software~\citep{Gorski_2005} in the simulations of our work. We work with the HEALPix resolution parameter $N_{\rm{side}}=256$. In this resolution, we produce the realizations of the full sky $Q$ and $U$ polarization maps of CMB, using \texttt{healpy.sphtfunc.synfast} with randomly chosen seed values, from the theoretical $E$- and $B$-modes spectra ($C_{\ell}^{th,EE}$ and $C_{\ell}^{th,BB}$) for the maximum multipole $\ell_{\rm{max}}=2N_{\rm{side}}=512$. We also apply the pixel window function ($P_{\ell}$) in \texttt{healpy.sphtfunc.synfast} for the estimation of these CMB polarization maps. In the left panel of the Figure~\ref{QU_map_full}, we show the full sky pixelated $Q$ map for a randomly selected seed value. In the right panel of this figure, we present the full sky pixelated $U$ map for the same seed value.
\subsection{Instrumental noise}
\label{noise_sec}
Depending upon the sensitivity of the telescope, the CMB polarization experiments observe the signal with a instrumental noise. Therefore, in our analysis, we use a detector noise level, i.e., baseline polarization map depth $\sigma_{d} = \rm{2.1 \ \mu K_{CMB}arcmin}$ which is compatible to $129$ GHz frequency band of the next generation satelite based CMB polarization experiment PICO~\citep{Hanany_2019}. The pixel variance of noise map is given by
\begin{eqnarray}
\sigma_{n}^{2} &=& \left(\frac{\sigma_{d}\pi}{180 \times 60}\right)^{2}\left(\frac{N_{\rm{pix}}}{4\pi}\right) \ , \label{noise_var}
\end{eqnarray}
where $N_{\rm{pix}}=12N_{\rm{side}}^{2}$. We generate the white noise realizations for $N_{\rm{side}}=256$. Moreover, the noise realizations of $Q$ polarization map are uncorrelated to the same of $U$ polarization map, since the seed values used for obtaining noise realization from the Gaussian distribution (with $0$ mean and $\sigma_{n}$ standard deviation) are different for each of these polarizations. The incorporation of these instrumental noises in partial sky polarization maps is discussed in section~\ref{part_QU_map_sec}.
\begin{figure*}[h!]
\centering
\includegraphics[scale=0.48]{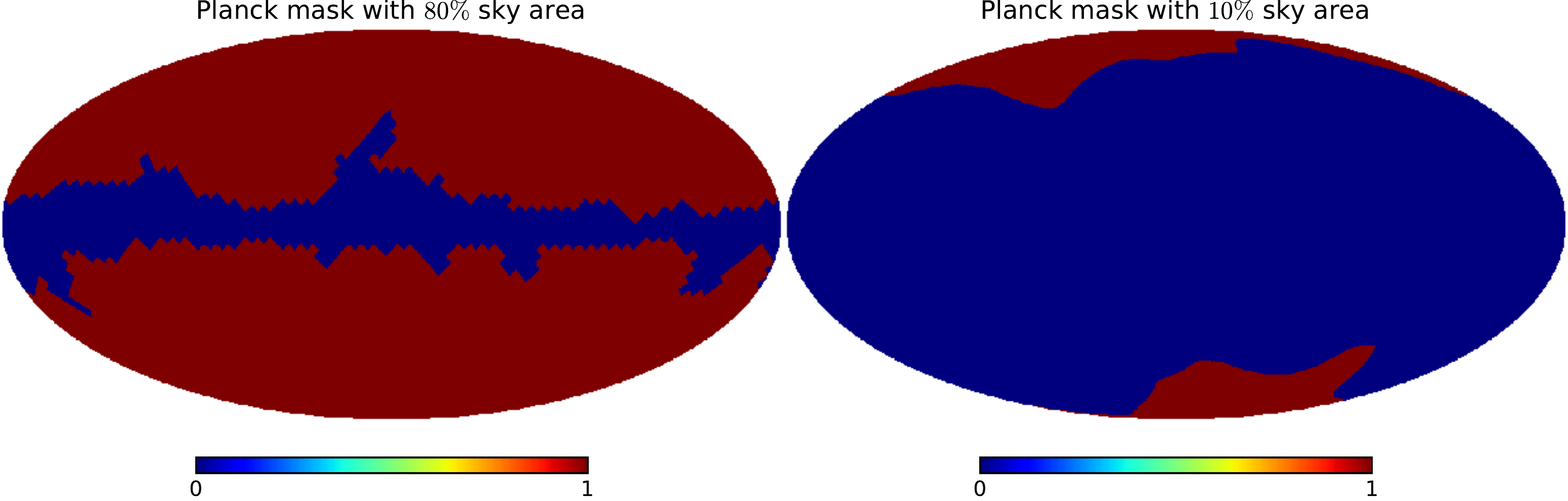}
\caption{Left panel shows the mollweide projection of the binary mask, namely Mask80, containing about $80\%$ sky area and right panel presents the mollweide projection of the binary mask, namely Mask10, containing about $10\%$ sky area for the resolution parameter $N_{\rm{side}}=256$. We refer to the section~\ref{part_QU_map_sec} for the detailed descriptions about the construction of these binary masks.}
\label{pmask}
\end{figure*}
\begin{figure*}[h!]
\centering
\includegraphics[scale=0.55]{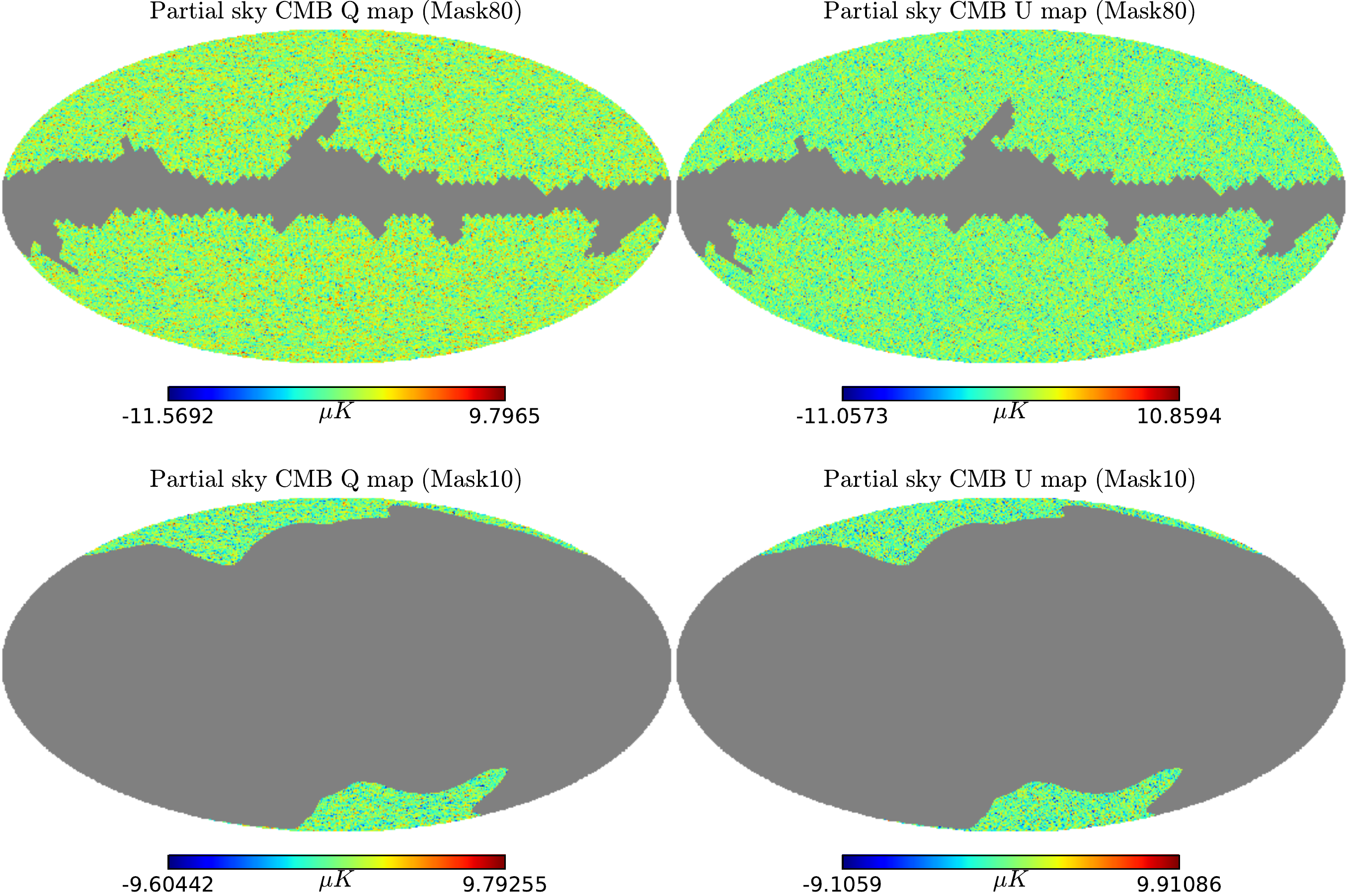}
\caption{Top left panel shows the partial sky $Q$ map and top right panel represents the partial sky $U$ map using Mask80 at the resolution $N_{\rm{side}}=256$ for a randomly chosen seed value. Bottom left panel presents the partial sky $Q$ map and bottom right panel shows the partial sky $U$ map using Mask10 at the same resolution for the same seed value. These partial sky maps are contaminated by detector noises which are uncorrelated for $Q$ and $U$ polarizations. We use gray colour in the masked region. For the unmasked region, the colour bar of each panel of the figure is in $\mu K$ unit.}
\label{QU_map_part}
\end{figure*}
\subsection{Partial sky $Q$ and $U$ maps of CMB polarization}
\label{part_QU_map_sec}
We consider two cases in our analysis using two binary masks to produce partial sky maps in each of these two cases. We denote these two binary masks by the names of Mask80 and Mask10, where Mask80 contains about $80\%$ of the sky area and Mask10 holds approximately $10\%$ sky area. We note that Mask10 is suitable for the typical ground based observations. These binary masks are created from the P-mask\footnote{\url{https://pla.esac.esa.int/\#results}} and the large galactic mask\footnote{\url{https://pla.esac.esa.int/\#results}} provided by~\cite{PlanckIV_2020}. Moreover, we incorporate the suitable detector noise(see section~\ref{noise_sec}) in the polarization maps smoothed by FWHM $7.4$ arcmin~\citep{Hanany_2019}. Then, we remove the beam effect from the noise contaminated and smoothed full sky maps before applying the binary mask to produce partial sky for each of these two cases. In the next two paragraphs, we describe the procedure to create Mask80 and Mask10 respectively.

P-mask is available in Planck's website with the resolution $N_{\rm{side}}=2048$. Atfirst, we downgrade this mask at the resolution $N_{\rm{side}}=16$ using \texttt{healpy.pixelfunc.ud\_grade}. Then, we convert this $N_{\rm{side}}=16$ mask to a binary mask assigning all the pixel values larger than $0.5$ to the new value of unity and the rest of the pixel values to zero. Moreover, we remove the point-sources masked regions, assigning to the value of unity, from this binary mask. Afterthat, we produce $N_{\rm{side}}=256$ binary mask (without any point-sources regions) upgrading the binary mask ($N_{\rm{side}}=16$) at the pixel resolution $N_{\rm{side}}=256$, using \texttt{healpy.pixelfunc.ud\_grade}. This final binary mask containing approximately $80\%$ sky area is named Mask80 for $N_{\rm{side}}=256$, which is shown in the left panel of Figure~\ref{pmask}.

The large galactic mask is available in Planck's website with 5 degree apodization and containing $20\%$ sky area for the resolution $N_{\rm{side}}=2048$. Atfirst, we downgrade this mask at the resolution $N_{\rm{side}}=256$ using \texttt{healpy.pixelfunc.ud\_grade}. Then, we convert this $N_{\rm{side}}=256$ mask to a binary mask assigning all the pixel values larger than $0.6$ to the new value of unity and the rest of the pixel values to zero. This binary mask namely Mask10 shown in the right panel of the Figure~\ref{pmask}, for $N_{\rm{side}}=256$, contains only $10\%$ sky area approximately.

Finally, we generate the partial sky $Q$ and $U$ maps for $N_{\rm{side}}=256$ and $\ell_{\rm{max}}=512$ by applying the binary mask (i.e., Mask80 or Mask10) on the noise contaminated full sky $Q$ and $U$ maps (Figure~\ref{QU_map_full}) after removing the smoothing effect from these full sky maps. In the top panels of Figure~\ref{QU_map_part}, we present the partial sky $Q$ and $U$ maps corresponding to Mask80 for a randomly chosen seed value and the resolution parameter $N_{\rm{side}}=256$. In the bottom panels of the Figure~\ref{QU_map_part}, we show the partial sky $Q$ and $U$ maps corresponding to Mask10 for the same seed value and the same resolution parameter.
\subsection{Simulations of full sky spectra of $E$- and $B$-modes}
\label{full_spec_sec}
\begin{figure}[h!]
\centering
\includegraphics[scale=0.34]{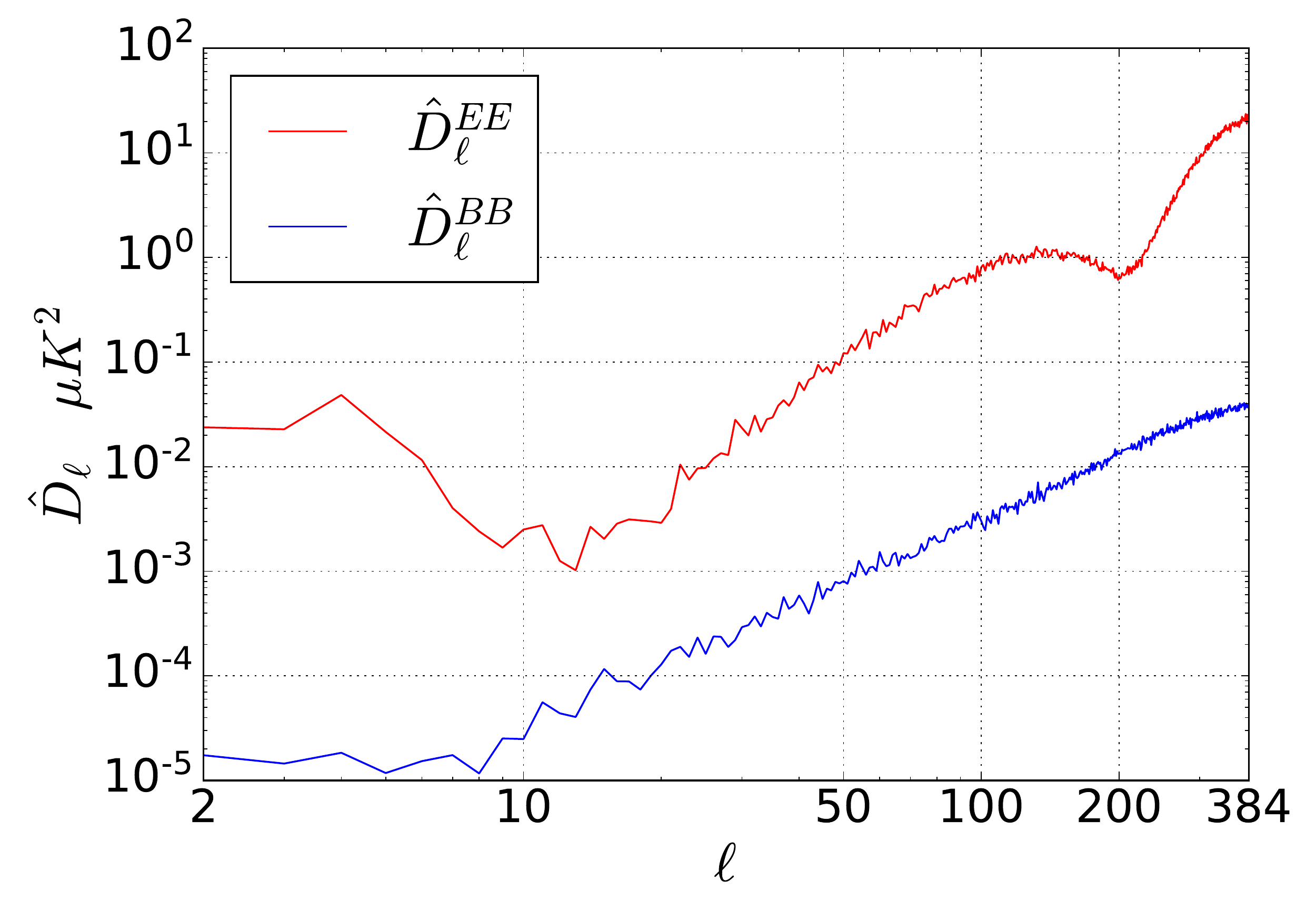}
\caption{Figure shows the full sky target power spectra for a randomly selected seed value, for the multipole range $2 \leq \ell \leq 384$, corresponding to $E$- and $B$- modes obtained from the full sky $Q$ and $U$ maps shown in Figure~\ref{QU_map_full}, where $\hat{D}_{\ell}=\ell(\ell +1)\hat{C}_{\ell}/2\pi$ in $\mu K^2$ unit and the `hat' notation define the target spectra. Both axes of the figure are in $\log_{10}$ scale.}
\label{full_sky_realization}
\end{figure}
\begin{figure*}[h!]
\centering
\includegraphics[scale=0.4]{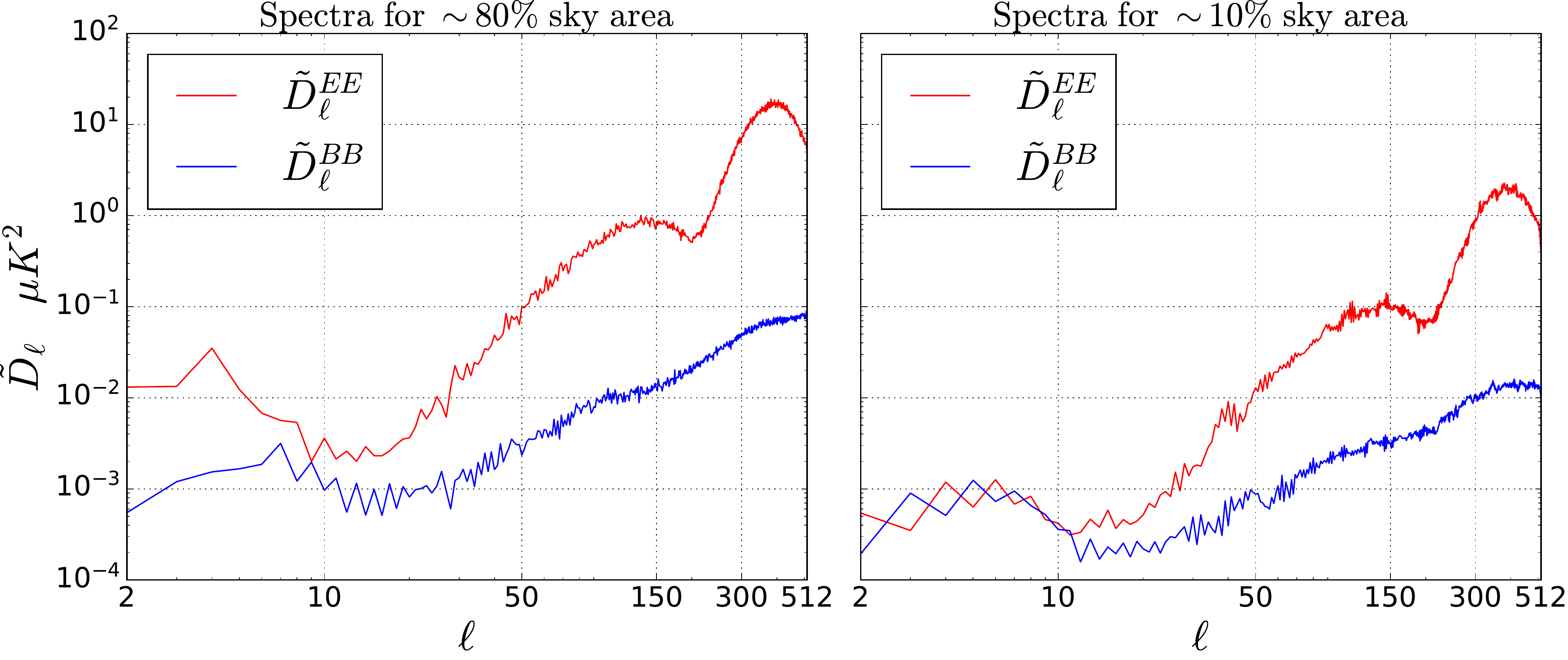}
\caption{Left panel represents the partial sky $E$- and $B$-modes power spectra corresponding to Mask80 and the right panel shows the partial sky spectra of these polarization modes corresponding to Mask10, for the multipole range $2 \leq \ell \leq 512$ for a randomly chosen seed value. These partial sky spectra are generated from the partial sky maps containing detector noise. Both axes of each panel of the figure are in $\log_{10}$ scale and $\tilde{D}_{\ell}$ equals to $\ell(\ell +1)\tilde{C}_{\ell}/2\pi$ in $\mu K^2$ unit. Comparing the each panel of this figure with Figure~\ref{full_sky_realization}, we conclude that the partial sky realization of $B$-mode spectra shows the significant power increase due to $E$-to-$B$ mode leakage.}
\label{part_sky_realization}
\end{figure*}
We use the full sky $Q$ and $U$ CMB polarization maps in \texttt{healpy.sphtfunc.anafast}, applying ring-weighting and using the maximum multipole $\ell_{\rm{max}}=384$, to generate the pixel-smoothed full sky power spectra ($\hat{C}_{\ell}^{pix,EE}$ and $\hat{C}_{\ell}^{pix,BB}$) corresponding to the CMB $E$- and $B$-modes. We utilize the `hat' notation to define the target spectra. We use $\ell_{\rm{max}}=1.5N_{\rm{side}}=384$ for producing these full sky spectra, since the estimation of power spectra by \texttt{healpy.sphtfunc.anafast} are more accurately to machine precision for the band-width limited input signal with $\ell_{\rm{max}} \leq 1.5N_{\rm{side}}$. We obtain the full sky $E$- and $B$-modes spectra ($\hat{C}_{\ell}^{EE}$ and $\hat{C}_{\ell}^{BB}$) from the pixel-smoothed spectra (corresponding to these CMB polarization modes) divided by the square of the pixel window function ($P_{\ell}$). In Figure~\ref{full_sky_realization}, we show the full sky target spectra ($\hat{D}_{\ell}^{EE}$ and $\hat{D}_{\ell}^{BB}$) corresponding to the full sky $Q$ and $U$ maps (Figure~\ref{QU_map_full}) for a randomly chosen seed value for the multipole range $2 \leq \ell \leq 384$, where $\hat{D}_{\ell} = \ell(\ell +1)\hat{C}_{\ell}/2\pi$ in $\mu K^2$. We produce $1.2 \times 10^5$ number of realizations of the full sky $E$- and $B$-modes spectra for our analysis. Ensemble average of these realization spectra of each polarization mode agrees with the corresponding theoretical spectrum.
\subsection{Simulations of partial sky spectra of $E$- and $B$-modes}
\label{part_spec_sec}
We obtain the impure partial sky $E$- and $B$-mode spectra ($\tilde{C}_{\ell}^{EE}$ and $\tilde{C}_{\ell}^{BB}$) by using the noise contaminated partial sky $Q$ and $U$ maps (as input maps) in \texttt{healpy.sphtfunc.anafast} with ring-weighting and the maximum multipole $\ell_{\rm{max}}=512$. We apply $\ell_{\rm{max}}=2N_{\rm{side}}=512$ for obtaining these partial sky spectra, since the use of $\ell_{\rm{max}} \approx 2N_{\rm{side}}$ is sufficient for the estimation of power spectrum with good accuracy. Another reason to use of $\ell_{\rm{max}}=512$ is that it helps to provide more information in the input of our CNN system (discussed in section~\ref{cnn_sec}) for predicting full sky spectra from the partial sky spectra of CMB polarization. In the left panel of Figure~\ref{part_sky_realization}, we show the realization of partial sky spectra ($\tilde{D}_{\ell}^{EE}$ and $\tilde{D}_{\ell}^{BB}$) corresponding to partial sky $Q$ and $U$ maps containing approximately $80\%$ sky area (shown in the top panels of Figure~\ref{QU_map_part}) with the multipole range $2 \leq \ell \leq 512$ for a randomly selected seed value, where $\tilde{D}_{\ell}=\ell(\ell +1)\tilde{C}_{\ell}/2\pi$. Similarly, in the right panel of Figure~\ref{part_sky_realization}, we present the realization of partial sky spectra ($\tilde{D}_{\ell}^{EE}$ and $\tilde{D}_{\ell}^{BB}$) obtained from partial sky $Q$ and $U$ maps containing about $10\%$ sky area (shown in the bottom panels of Figure~\ref{QU_map_part}) for the same multipole range and same seed value. We note in passing that there is a significant power increase in the partial sky spectra of $B$-mode (comparing between Figure~\ref{full_sky_realization} and each panel of Figure~\ref{part_sky_realization}), since the leakage between $E$- and $B$-modes of CMB polarization appears in the partial sky analysis. We also generate $1.2 \times 10^5$ number of realizations of the partial sky $E$- and $B$-modes spectra for the CNN analysis in case of each of two different masks.
\subsection{CNN for our analysis}
\label{cnn_sec}
\begin{figure*}[h!]
\centering
\includegraphics[scale=0.4]{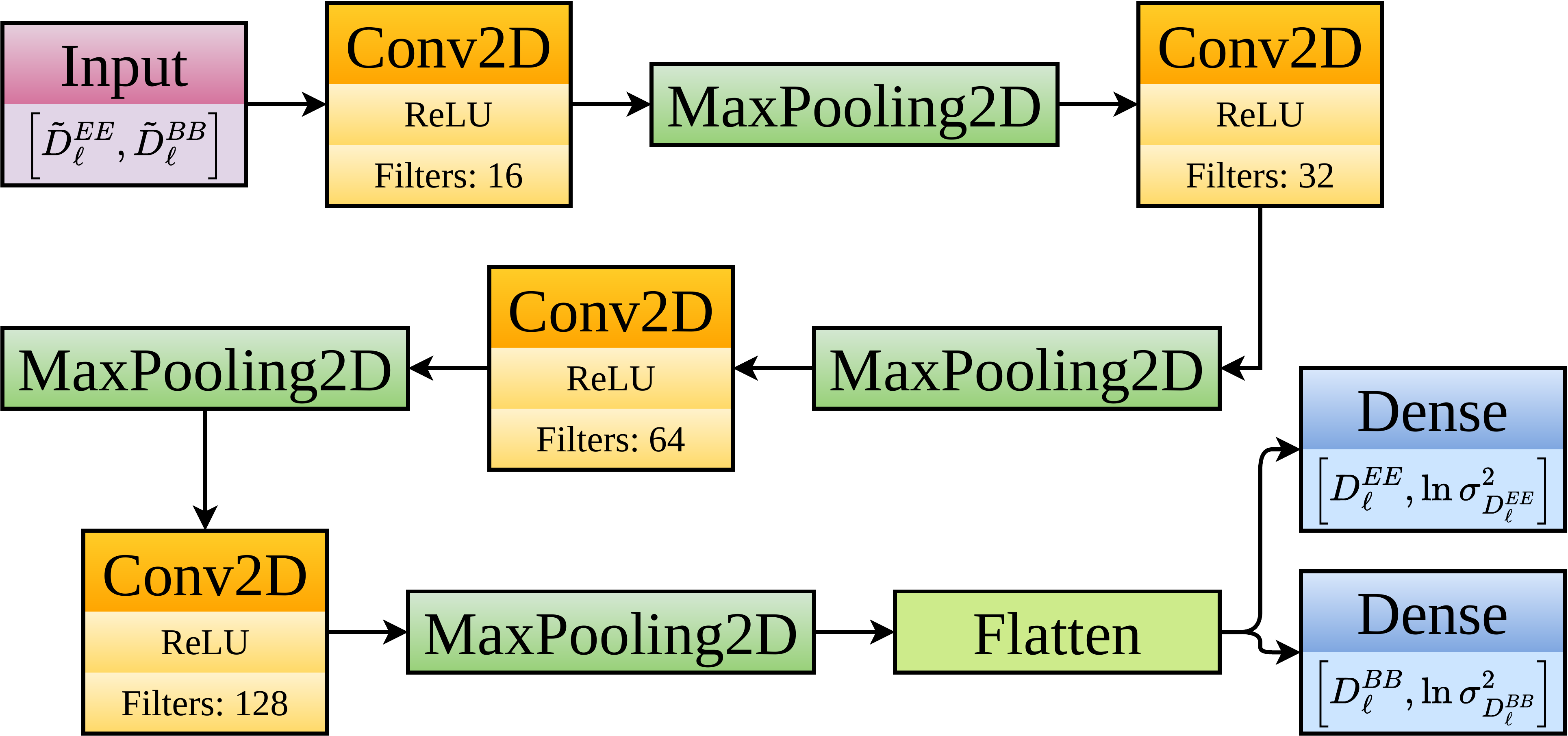}
\caption{Figure shows the flowchart of our CNN system. We present the non-linear activation function (ReLU) and the number of filters used in the hidden (Conv2D) layers. We employ MaxPooling2D after each hidden layer and also use the Flatten layer before the output layers. We use the partial sky spectra ($\tilde{D}_{\ell}^{EE}$ and $\tilde{D}_{\ell}^{BB}$) in input layer for the multipole range $2 \leq \ell \leq 512$ and predict the full sky spectra ($D_{\ell}^{EE}$ and $D_{\ell}^{BB}$) with the log variances ($\ln\sigma_{D_{\ell}^{EE}}^{2}$ and $\ln\sigma_{D_{\ell}^{BB}}^{2}$) corresponding to these spectra for the multipole range $2 \leq \ell \leq 384$ in two Dense output layers. We refer the section~\ref{cnn_sec} for the detailed discussion about our CNN system.}
\label{CNN_model}
\end{figure*}
We use TensorFlow\footnote{\url{https://www.tensorflow.org/}}~\citep{Abadi_2015} ML platform to organize the CNN system as well as to train this CNN system for the unbiased predictions of the full sky spectra of $E$- and $B$-modes from the partial sky spectra of these modes.

In deep learning, CNNs are the special types of the feed-forward neural network. They can contain multiple hidden layers in between input and output layers, like traditional ANNs. In traditional ANNs, input features are used as a one dimensional array. However, we can directly provide the multi-dimensional input (e.g., image, audio, video) in CNNs without breaking the pixelized patterns of the input. This is one of the major advantages of the using of CNNs. We refer these literatures (i.e.,~\cite{Oshea_2015,Gu_2015}) for the detailed discussions of the basics of CNNs and its applications in the various fields.

In our analysis, we create a CNN architecture for the supervised-learning of the mapping from the partial sky $E$- and $B$-modes power spectra (for the multipole range $2 \leq \ell \leq 512$) to the full sky spectra of these modes (for the multipole range $2 \leq \ell \leq 384$). In Figure~\ref{CNN_model}, we show the flowchart of our CNN system. We use the partial sky spectra (input ; $\bigl[\tilde{D}_{\ell}^{EE}$, $\tilde{D}_{\ell}^{BB}\bigr]$) of $E$- and $B$-modes, which are labelled by the full sky spectra (target ; $\bigl[\hat{D}_{\ell}^{EE}$, $\hat{D}_{\ell}^{BB}\bigr]$, where the `hat' notation is used to define the target spectra) of these modes, as the input features in the input layer of the CNN. These input features are stored in a matrix with the shape ($2,511$), where the former row contains the $E$-mode spectrum and the latter is for the $B$-mode spectrum corresponding to the partial sky. Then, we provide this input matrix with grey colour scale in the input layer. Consequently, this input matrix (with grey colour scale) is carried out from the input layer to the first Conv2D layer. We use the kernel of shape ($1,5$) for the convolutional operation and also utilize \texttt{L2} kernel regularizer (with the hyperparameter value $0.01$) to avoid any types of overfitting in each Conv2D layer. We also use the ReLU activation function in each Conv2D layer to learn the non-linear relation between input and target data. We employ the MaxPooling2D after each Conv2D to downgrade the output shape of these Conv2D layers. We use the window shape ($1,4$) in the top three MaxPooling2D in our CNN system and the last one operates with the window size ($2,2$). To preserve all the information of the input data, we set the number of the filters in four hidden (Conv2D) layers as $16,32,64,128$ respectively, since the output shapes in these hidden layers are downgraded by MaxPooling2D in lower size along the forward direction of our CNN system. We use the Flatten layer to compress the output of the previous MaxPooling2D layer in a single array. We use two Dense layers with linear activation function for predicting the full sky spectra. First half of the output of each output layer contains the predicted values of the full sky polarization spectrum for the multipole range $2 \leq \ell \leq 384$ and the rest half takes care of the log variances corresponding to these predicted values for the same multipole range, since we use the heteroscedastic loss (HS) function~\citep{Kendall_2017} in both output layers to train our CNN system. This HS loss function can be defined as
\begin{eqnarray}
L^{\rm{HS}} \ = \ \frac{1}{2n}\sum\limits_{q=0}^{n}\left[\exp\left(-s_{q}\right)\left(y_{q}-\hat{y_{q}}\right)^{2}+s_{q}\right] \ , \label{L_hs}
\end{eqnarray}
where $s_{q}$ is the $q$-th log variances $\ln\left(\sigma_{q}^{2}\right)$ and $\sigma_{q}$ is the aleatoric uncertainties~\citep{Kendall_2017} corresponding to the prediction values ($y_{q}$). Moreover, $n$ defines the number of the target values ($\hat{y}_{q}$) in each output layer, which is $383$. We particularly use this HS loss function to calculate the aleatoric uncertainties. These uncertainties are produced due to the inherent noise in the input data, since these input partial sky spectra are obtained by employing the mask in the full sky spectra.

Firstly, we randomly shuffle the total $1.2 \times 10^5$ number of simulated realizations of the partial and full sky spectra corresponding to both $E$- and $B$-modes CMB polarization. After random shuffling, we use the first $10^5$ number of samples of input and target data for the training process of our CNN system. From the rest of the samples, we use the first half for the validation and the second half for the testing of our CNN system. We apply the standardization method for the preprocessing of the training samples of the input data. For this standardization method, firstly we calculate the mean and the standard deviation of the training samples in case of input features. Then, we divide each input realization by the input standard deviation after subtracting by the input mean at each multipole. We also apply these same mean and standard deviation of the training samples to standardize the input samples used in the validation process. Moreover, the input mean and standard deviation of the training samples are also used to standardize the partial sky spectra of test set for the predictions of the corresponding full sky spectra. We also scale the target full sky spectra by taking $\log_e$ of these target spectra and then dividing by 10 for training and validation sets. We assign the batch size equals to $1024$ and epochs equals to 100 at the time of fitting our CNN system with the simulated data to utilize the mini-batch algorithm in our CNN system. Considering the special type of loss function (equation~\ref{L_hs}) as well as the mini-batch algorithm, we optimize our CNN system using Adam optimizer~\citep{Kingma_2014} with the learning rate value $10^{-4}$.

We utilize the model-averaging-ensemble method~\citep{Lai_2021}, which is a computationally low cost process, to reduce the epistemic uncertainties of our CNN system. For this method, we concentrate on the random initializations of the convolutional kernels in Conv2D layers. We train our CNN system with the same hyperparameter values for a total of $100$ times for $100$ randomly selected seed values using TensorFlow library. We use $\texttt{Google Colab}$\footnote{\url{https://colab.research.google.com/}}, a ML platform for an efficient online GPU service offered by Google, for the entire training process of our CNN system. It takes approximately 7 hours to perform this entire training process. We find the predictions ($y_{q}^{\rm{mean}}$) by taking the simple mean of these $100$ output sets. For the estimation of the aleatoric uncertainties ($\sigma_{q}^{\rm{rms}}$; $\rm{rms}$ stands for root mean square) corresponding to these predictions, firstly we calculate the exponential of the log variances ($s_{q}$) for the same output sets. Afterthat, we estimate the simple mean of these $\exp(s_{q})$ from these output sets. Then, the aleatoric uncertainties ($\sigma_{q}^{\rm{rms}}$) are obtained by taking the square root of the mean values of $\exp(s_{q})$. Thereafter, we estimate the predictions by computing the sum of the mean predictions ($y_{q}^{\rm{mean}}$) and the random Gaussian realizations created from the Gaussian distributions with $0$ mean and $\sigma_{q}^{\rm{rms}}$ standard deviation. Finally, as we also scale the target spectra for the training process, we calculate the final predictions of the full sky $E$- and $B$-modes spectra by taking exponential of the predictions of the test set after multiplying by 10.
\section{Results}
\label{results}
\subsection{Predicted realization spectra of $E$- and $B$-modes}
\label{realization_sec}
\begin{figure*}[h!]
\centering
\includegraphics[scale=0.4]{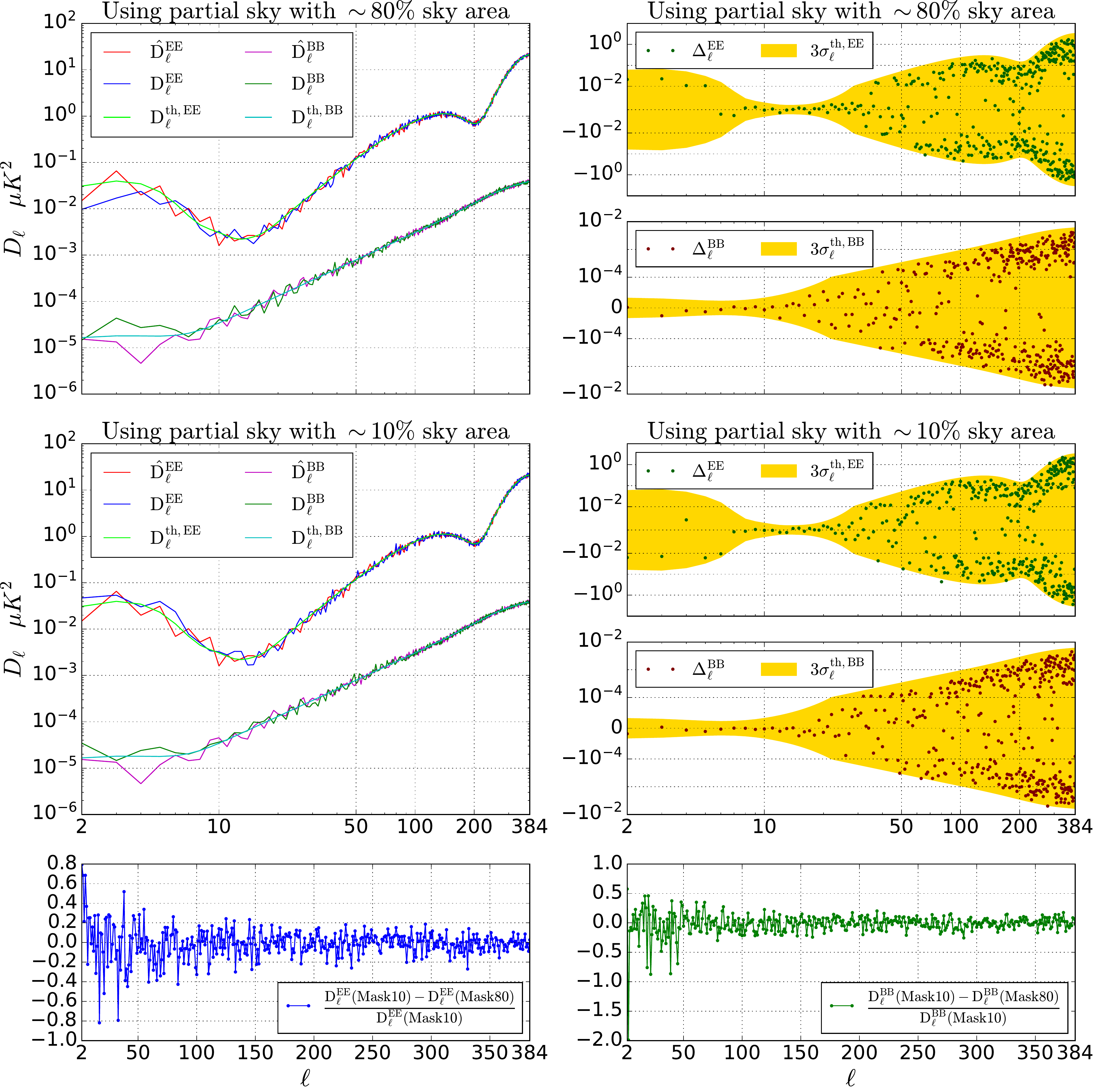}
\caption{Top left sub-figure shows the agreement between the target and the predicted spectra of the CMB polarization modes for the multipole range $2 \leq \ell \leq 384$ for a randomly chosen test sample corresponding to Mask80. Middle left sub-figure shows the same plot as the top left sub-figure but corresponding to Mask10. First two sub-figures of right column present the differences between the theoretical and predicted spectra for $E$ and $B$-modes respectively for Mask80. Similarly, third and fourth sub-figures of the right column show the differences between the theoretical and the predicted spectra of same modes corresponding to Mask10. In these four sub-figures of the right column, horizontal and vertical axes are presented in log and semi-log (e.g., region near zero value is in linear scale) scales respectively. Consequently, these four sub-figures of the right column show that the most of these differences are within three times cosmic standard deviation (yellow) for these polarization spectra in case of each mask. In the bottommost row, we show the relative difference between spectra of Mask10 and Mask80 at each multipole for $E$-mode in bottom left and for $B$-mode in bottom right.}
\label{realization_EB}
\end{figure*}
\begin{figure*}[h!]
\centering
\includegraphics[scale=0.4]{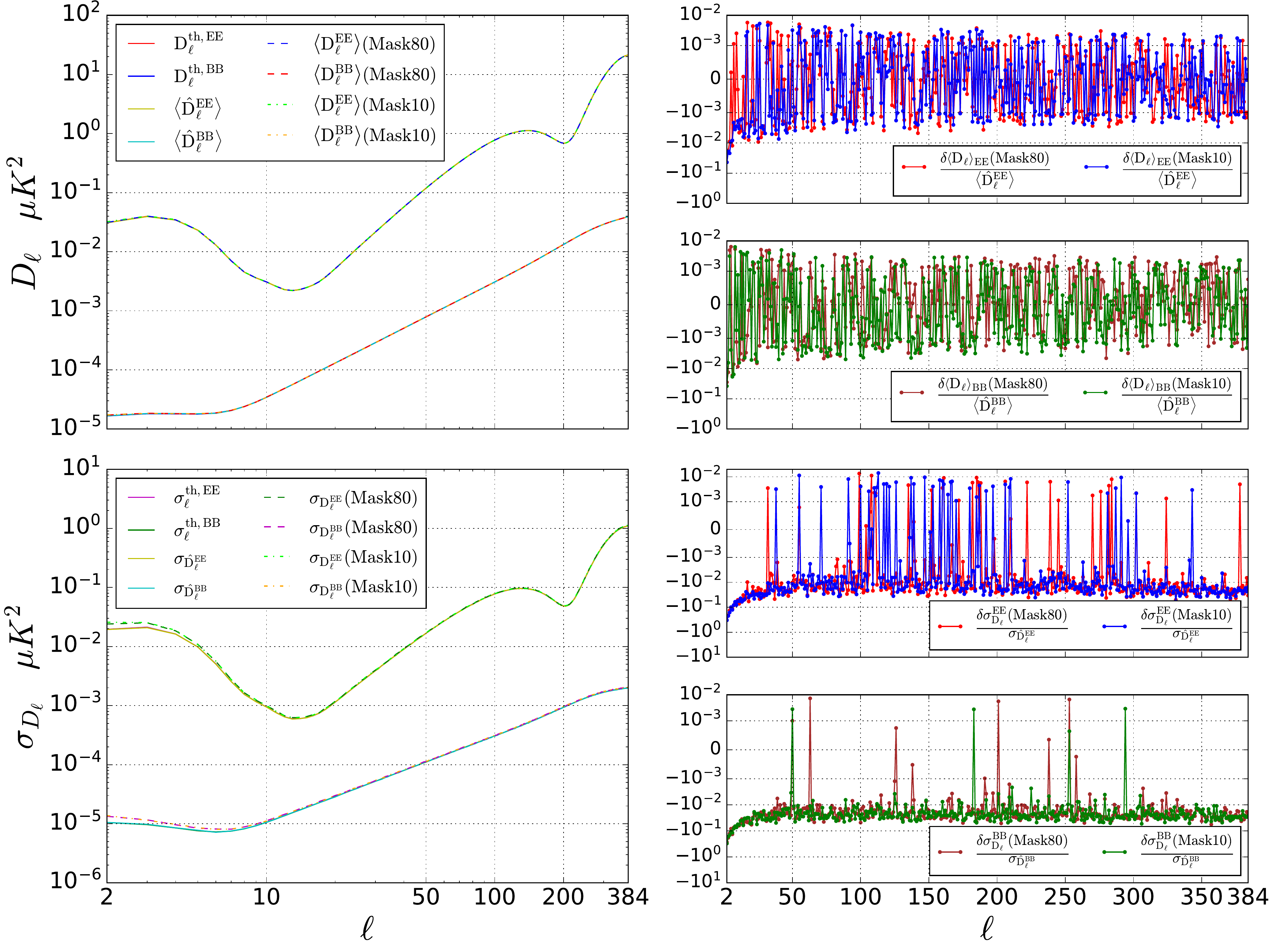}
\caption{Top sub-figure of the left column shows the agreement of the mean predicted spectrum with the mean target spectrum and also with the theoretical spectrum for each of the CMB polarization modes (i.e., $E$- and $B$-modes) for both masks. Similarly, in the bottom sub-figure of the left column, the standard deviation of the predicted spectra shows the agreement with the standard deviation of the target spectra as well as the cosmic standard deviation for each of these polarization modes for both masks. In the left column, we use log scale along both axes in each sub-figure. In the right column of the figure, first two sub-figures show the relative differences between the mean target and the mean predicted spectrum, and last two sub-figures present the relative differences between the standard deviations of the target and predicted spectra for $E$- and $B$-modes polarization for both masks. We use semi log scale along vertical axes in the sub-figures of right column. We refer section~\ref{mean_std_sec} for the detailed discussion.}
\label{mean_std}
\end{figure*}
First, we train our CNN system using the partial sky $E$- and $B$-modes power spectra obtained by employing Mask80. Using this trained CNN system, we predict $10^4$ number of realizations of the full sky power spectra ($D_{\ell}^{EE}$ and $D_{\ell}^{BB}$) corresponding to the CMB polarization $E$- and $B$-modes from the partial sky spectra ($\tilde{D}_{\ell}^{EE}$ and $\tilde{D}_{\ell}^{BB}$) of these two polarization modes. These predictions of the full sky spectra are in the multipole range $2 \leq \ell \leq 384$ for each of these polarization modes. In the top left sub-figure of Figure~\ref{realization_EB}, we present the predicted full sky realization spectra of these two polarization modes compared with the corresponding target spectra ($\hat{D}_{\ell}^{EE}$ and $\hat{D}_{\ell}^{BB}$) for a randomly chosen test samples. We note that the predicted realization spectra $D_{\ell}^{EE}$ and $D_{\ell}^{BB}$ agree excellently with the corresponding target realization spectra $\hat{D}_{\ell}^{EE}$ and $\hat{D}_{\ell}^{BB}$. Moreover, we estimate the differences between the theoretical and predicted spectra for each of these two polarization modes. These differences are given by
\begin{eqnarray}
\Delta_{\ell}^{x} &=& D_{\ell}^{th,x} - D_{\ell}^{x} \ , \label{delta_EB}
\end{eqnarray}
where $x$ indicates $EE$ or $BB$. In the first two sub-figures of the right column of the Figure~\ref{realization_EB}, we present the differences corresponding to $E$-mode spectra and the differences corresponding to $B$-mode spectra respectively. In these same sub-figures, we also show three times cosmic standard deviation region corresponding to these polarization spectra. We note that our CNN system predicts the full sky $E$- and $B$-modes spectra reliably, since the most of the differences of the spectra of each polarization mode are within the corresponding three times cosmic standard deviation region.

Secondly, we train our same CNN system using the partial sky $E$- and $B$-modes spectra obtained from partial sky maps corresponding to Mask10 and predict $10^4$ number of realizations of full sky spectra (for the multipole range $2 \leq \ell \leq 384$) from the partial sky spectra corresponding to this Mask10. In case of this large galactic mask, we also show the agreement between realization full sky spectra of $E$ and $B$-modes with the corresponding target spectra along with the theoretical spectra in the middle left sub-figure of Figure~\ref{realization_EB}. Following equation~\ref{delta_EB}, in two bottom sub-figures of the right column of Figure~\ref{realization_EB} we also present the differences between the theoretical and predicted spectra of $E$- and $B$-modes respectively along with three times cosmic standard deviations. We note that the full sky spectra of these polarization modes predicted by our CNN system are statistically reliable even in case of large galactic mask.
\begin{figure*}[h!]
\centering
\includegraphics[scale=0.45]{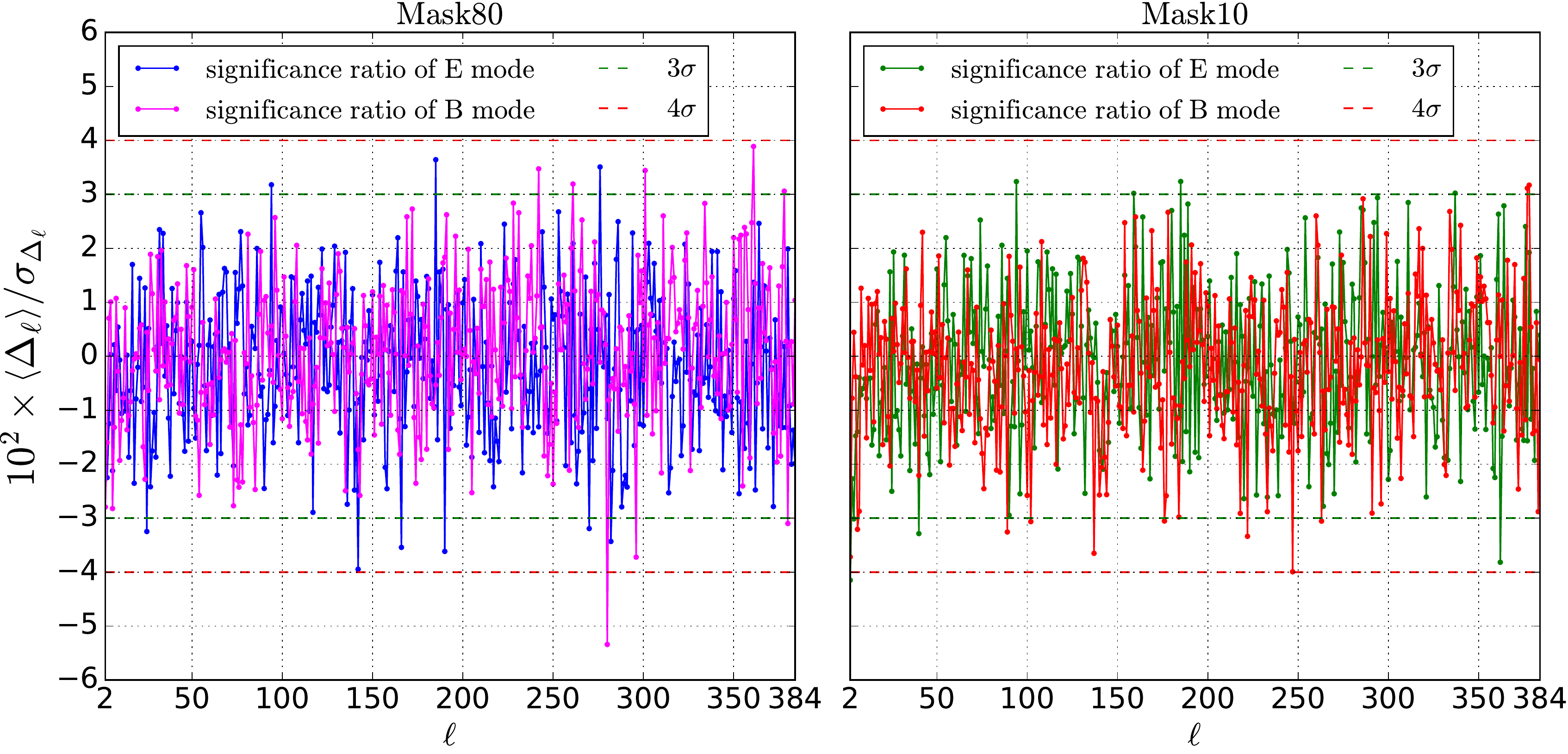}
\caption{Left panel: figure showing the significance ratios for the predicted full sky power spectra of $E$- and $B$-modes corresponding to Mask80. Right panel: figure representing the significance ratios for the predicted spectra of these polarization modes corresponding to Mask10. We also show the $3\sigma$ (green dashed line) and $4\sigma$ (red dashed line) error in each panel of the figure. We refer the section~\ref{sign_sec} for the detailed discussion.}
\label{sign}
\end{figure*}

In the bottom left panel of Figure~\ref{realization_EB}, we show the relative differences between target spectra for Mask10 and Mask80 for $E$-mode. The same relative differences for $B$-mode are shown in the bottom right panel of this figure.
\subsection{Predicted $D_{\ell}$ and $\sigma_{D_{\ell}}$ of $E$- and $B$-modes}
\label{mean_std_sec}
We estimate the mean of $10^4$ predicted realizations of the power spectrum for the multipole range $2 \leq \ell \leq 384$ for each of the CMB polarization modes $E$ and $B$. We further calculate the standard deviation of these realization spectra for each of these polarization modes. We calculate these mean and standard deviation of predicted full sky polarization spectra for the case of each of two different maks used in our analysis. In the top sub-figure of the left column of the Figure~\ref{mean_std}, we present the mean target (i.e., $\bigl<\hat{D}_{\ell}^{EE}\bigr>$, $\bigl<\hat{D}_{\ell}^{BB}\bigr>$), the mean predicted (i.e., $\bigl<D_{\ell}^{EE}\bigr>$, $\bigl<D_{\ell}^{BB}\bigr>$) and the theoretical (i.e., $\bigl<D_{\ell}^{th, EE}\bigr>$, $\bigl<D_{\ell}^{th, BB}\bigr>$) spectra of these polarization modes for both masks. The mean predicted $E$- and $B$-modes spectra provide excellent agreement with the mean target spectra of these polarization modes for each mask. In the bottom sub-figure of the left column of this figure, we show the standard deviations corresponding to the target (i.e., $\sigma_{D_{\ell}^{EE}}$, $\sigma_{D_{\ell}^{BB}}$) and the predicted spectra (i.e., $\sigma_{\hat{D}_{\ell}^{EE}}$, $\sigma_{\hat{D}_{\ell}^{BB}}$) as well as the cosmic standard deviation (i.e., $\sigma_{\ell}^{th, EE}$, $\sigma_{\ell}^{th, BB}$) for both masks. Interestingly, the standard deviations of these $10^4$ predicted spectra also agree excellently with the standard deviations of the corresponding target spectra for each of these polarization modes corresponding to each mask. In first two panels of the right column of Figure~\ref{mean_std}, we also show the differences between the mean target and the mean predicted spectra (i.e., $\delta\bigl<D_{\ell}\bigr>_{EE}$, $\delta\bigl<D_{\ell}\bigr>_{BB}$) at each multipole for both masks. Similarly, in last two panels of the right column of this figure, we also present the differences between the standard deviations corresponding to the target and predicted spectra (i.e., $\delta\sigma_{D_{\ell}}^{EE}$, $\delta\sigma_{D_{\ell}}^{BB}$) at each multipole for these polarization $E$- and $B$-modes respectively for both masks. These differences of the mean spectra as well as the differences of the standard deviations of these spectra show that the full sky $E$- and $B$-modes spectra predicted by our CNN system are effectively reliable and unbiased, since the mean predicted spectra of these polarization modes preserve their corresponding cosmic variances without affecting by the sample variances of the spectra even for the case of the partial sky contained very less sky area.
\subsection{Significance ratios}
\label{sign_sec}
\begin{figure*}[h!]
\centering
\includegraphics[scale=0.4]{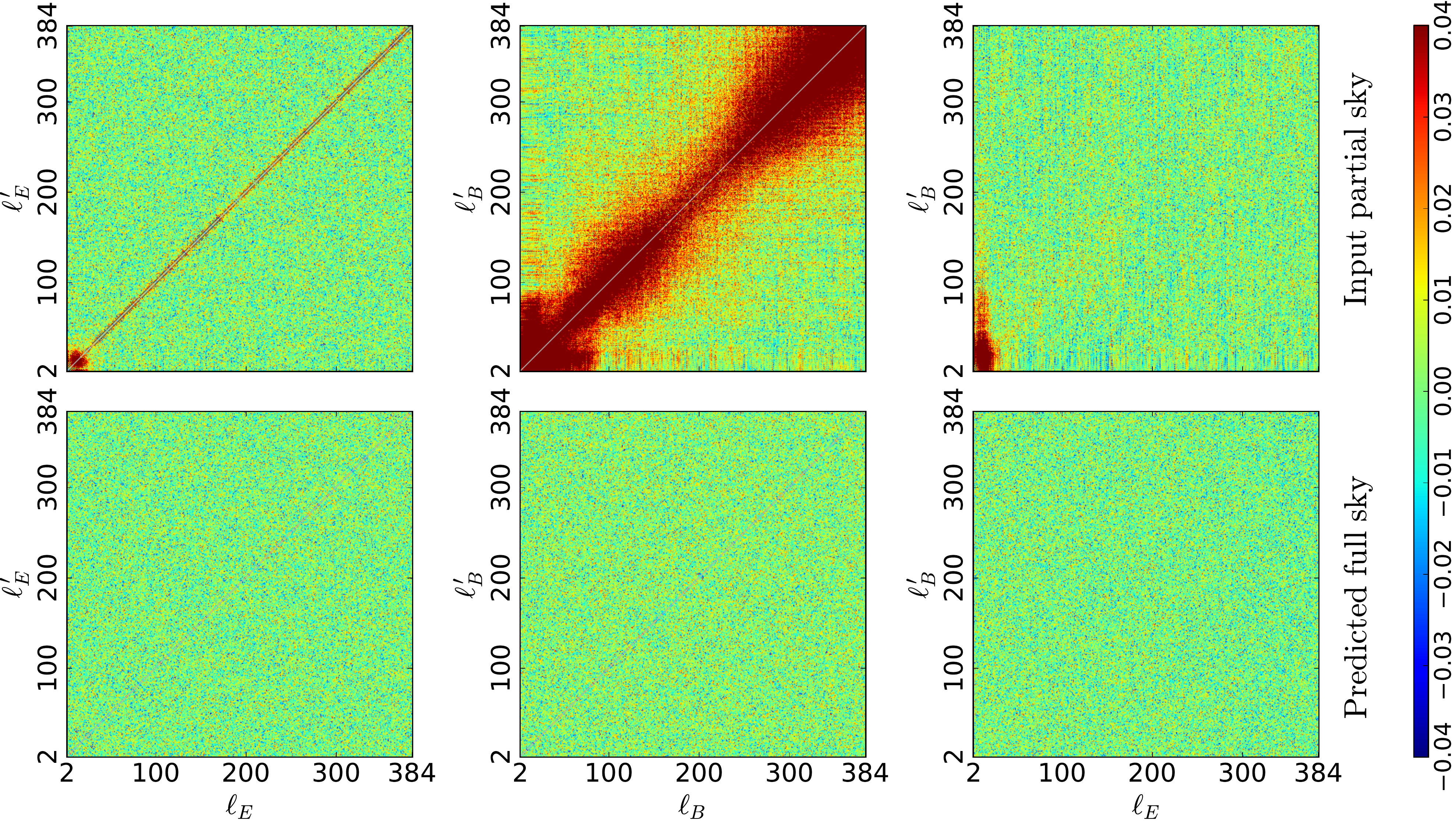}
\caption{Two rows of the figure show the multipole-multipole correlation matrices for the input partial sky and the predicted full sky spectra respectively for the multipole range $2 \leq \ell \leq 384$ for the partial sky analysis with Mask80. Moreover, the columns of the figure are corresponding to the correlations in the $E$-mode spectra, the correlations in the $B$-mode spectra and the correlations in between them respectively. We clip the colour bar within $\pm 0.04$. We also use white colour (in place of unity values) in diagonal of the matrices in the first two columns of the figure. These results shows that the correlations of each mode and the correlations between these two modes are excellently minimized in the predicted spectra. We refer the section~\ref{corr_sec} for the detailed discussion\protect\footnotemark.}
\label{corr_80}
\end{figure*}
We calculate the significance ratio at each multipole for each of the CMB polarization modes for both masks. In the left panel of Figure~\ref{sign}, we show the significance ratios of the predicted full sky spectra of these polarization modes corresponding to Mask80. In the right panel of Figure~\ref{sign}, we present the significance ratios of the predicted spectra of the same polarization modes but in case of the analysis using Mask10. In each panel of this figure, we also present the $3\sigma$ and $4\sigma$ error line using green and red dashed line respectively. Firstly, we calculate the differences between the target and predicted spectra for $10^4$ number of test samples. Then, we estimate the mean and standard deviation of these differences at each multipole. Thereafter, we obtain the standard error of mean (SEM) dividing the standard deviation by the square root of the number of test samples. Finally, we estimate the significance ratio as the mean difference divided by the SEM at each multipole. According to the Figure~\ref{sign}, we note that all the significance ratios are within the $4\sigma$ error line for each of these CMB polarization modes except one multipole point ($\ell=280$) in case of $\sim 80\%$ sky area for $B$-mode spectrum and one multipole point ($\ell=2$) in case of $\sim 10\%$ sky area for $E$-mode spectrum. More specifically, the significance ratios of the $E$-mode spectra are within $3\sigma$ error line except nine multipole points ($\ell$ $=$ $25,94,142,166,185,190,270,276,282$) and the significance ratios of the $B$-mode spectra are within the same error line except eight multipole points ($\ell$ $=$ $242,261,280,296,301,361,378,380$) for the partial sky analysis with Mask80. Similarly, we can state that the significance ratios of the $E$-mode spectra are within $3\sigma$ error line except eight multipole points ($\ell$ $=$ $2,4,40,94,159,185,337,362$) and the significance ratios of the $B$-mode spectra are within the same error line except eleven multipole points ($\ell$ $=$ $2,6,89,102,137,176,222,247,263,377,378$) for approximately $10\%$ sky analysis. These results also indicates the unbiased predictions of the full sky spectra of the CMB polarization modes by our CNN system in both cases.\footnotetext{We have seen that the correlation matrices for the target and predicted spectra contain similar values for the range of multipoles considered in this article.}
\subsection{Correlation matrix}
\label{corr_sec}
\begin{figure*}[h!]
\centering
\includegraphics[scale=0.4]{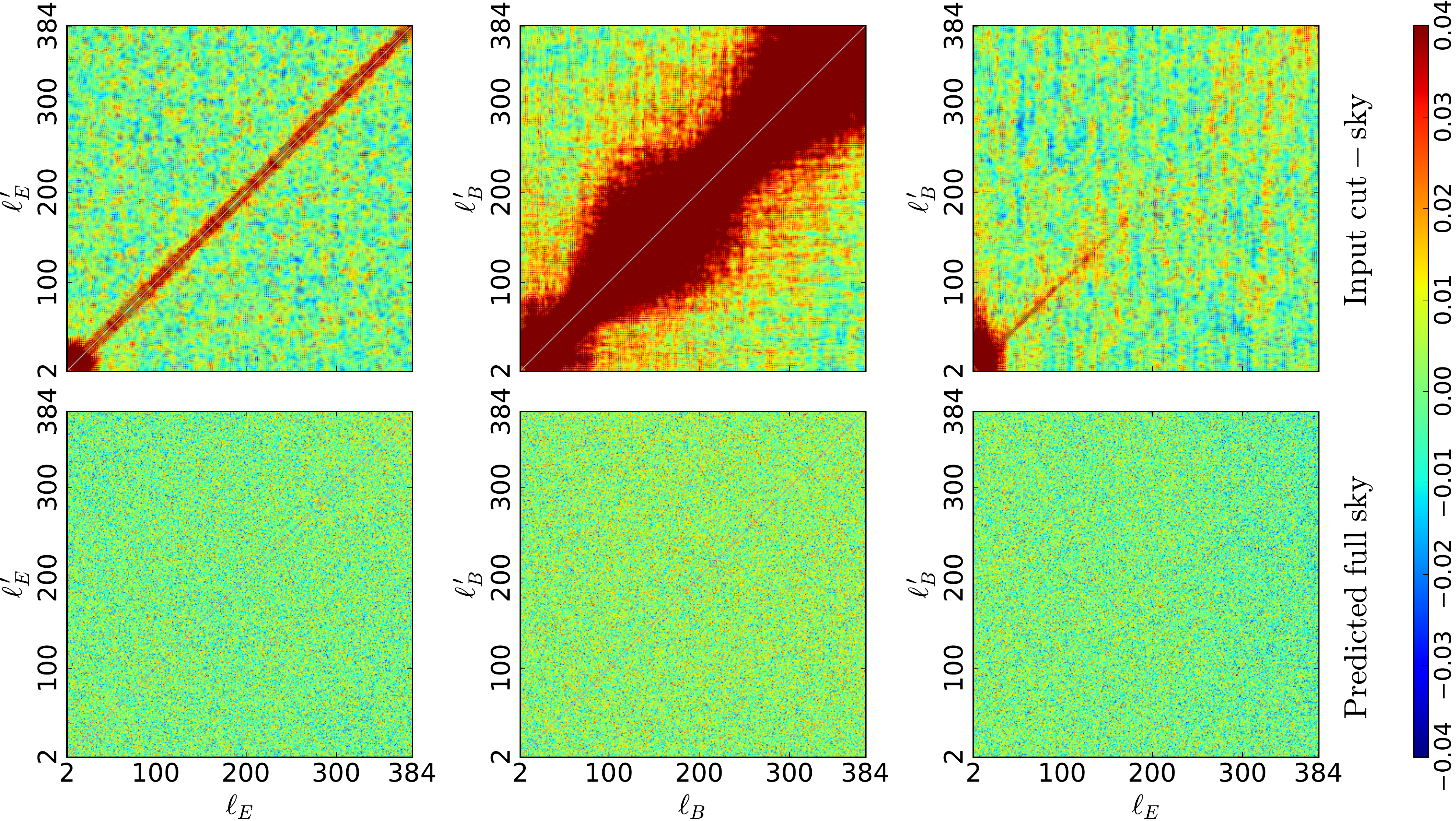}
\caption{Each panel of this figure presents the equivalent result shown in the same panel of Figure~\ref{corr_80} but for the partial sky analysis with Mask10. Here, we also clip the colour bar within $\pm 0.04$ and use white colour (in place of unity values) in diagonal of the matrices in the first two columns of the figure. We refer the section~\ref{corr_sec} for the detailed discussion\protect\footnotemark.}
\label{corr_10}
\end{figure*}
We calculate the multipole-multipole correlations in the predicted realizations of the $E$-mode spectrum, in the predicted realizations of the $B$-mode spectra, and also in between these polarization modes for the multipole range $2 \leq \ell \leq 384$ for each of Mask80 and Mask10. The correlation formula between the power spectra (i.e., $D_{\ell}$ and $D_{\ell'}$) at the multipoles $\ell$ and $\ell'$ is given by
\begin{eqnarray}
R_{\ell\ell'} &=& \frac{\sum\limits_{k}\left(D_{\ell}^{k}-\left<D_{\ell}\right>\right)\left(D_{\ell'}^{k}-\left<D_{\ell'}\right>\right)}{\sqrt{\sum\limits_{k}\left(D_{\ell}^{k}-\left<D_{\ell}\right>\right)^2\sum\limits_{k}\left(D_{\ell'}^{k}-\left<D_{\ell'}\right>\right)^2}} \ , \label{corr_form}
\end{eqnarray}
\footnotetext{Similar to Mask80, for Mask10 we have also seen that the correlation matrices for the target and predicted spectra contain similar values for the range of multipoles considered in this article.}where $\left<\cdots\right>$ indicates the simple mean of the samples and superscript $k$ denotes the $k$-th sample. We utilize this equation~\ref{corr_form} to estimate these correlations in the full sky spectra predicted by our CNN system. Using the same equation, we also calculate the correlations in the input cut-sky spectra for each of these polarization modes for the multipole range $2 \leq \ell \leq 384$. In equation~\ref{corr_form}, we use each of $D_{\ell}^{k}$, $D_{\ell'}^{k}$, $\left<D_{\ell}\right>$ and $\left<D_{\ell'}\right>$ for only $E$-mode to calculate the correlations in the realizations of the $E$-mode spectrum. We also follow the similar procedure to estimate the correlations in the realizations of $B$-mode spectrum using these variables of this equation for only $B$-mode. Nevertheless, we estimate the correlations between $E$- and $B$-modes spectra using $D_{\ell}^{k}$, $\left<D_{\ell}\right>$ for only $E$-mode and $D_{\ell'}^{k}$, $\left<D_{\ell'}\right>$ for only $B$-mode. In Figure~\ref{corr_80}, we present these correlation matrices corresponding to the input cut-sky as well as the predicted full sky spectra in case of about $80\%$ sky analysis. In Figure~\ref{corr_10} we show the correlations appeared in input cut-sky spectra as well as in the predicted full sky spectra of both polarization modes for approximately $10\%$ sky case. To understand the correlation matrices clearly, we clip the colour bar within $\pm0.04$ and use white colour in the place of unity diagonal values of these correlation matrices in both figures. In the left panel of the top row of each of these figures, we show the multipole-multipole correlations  in the realizations of the $E$-mode spectra for the input cut-sky (e.g., partial sky spectra corresponding to Mask80 or Mask10). These correlation matrices show the significant correlations in the low multipole region and along the diagonal region due to significant sky cutting (including the galactic region) of $Q$ and $U$ maps. The correlations of partial sky $E$-mode spectra are in between $-0.04$ and $0.243$ for Mask80, and in case of Mask10 it is within $-0.043$ and $0.976$. Due to the same reason, the significant correlations also appear in the correlation matrix corresponding to the $B$-mode spectra which is presented in the middle panel of the top row of each of these figures. The correlations of partial sky $B$-mode spectra are within $-0.037$ and $0.871$ for Mask80, and for the partial sky analysis with Mask10 the correlations of partial sky $B$-mode spectra are in between $-0.033$ and $0.954$. In the right panel of the top row of each of these figures, we present the correlations between the $E$- and $B$-modes spectra in the multipole spaces. Since there is the significant leakage of the power from the partial sky $E$-mode spectra to the partial sky $B$-mode spectra, this correlation matrix also shows the significant correlations between the partial sky spectra of these two CMB polarization modes in each of two different partial sky cases. The range of the correlations between partial sky $E$- and $B$-modes spectra is from $-0.043$ to $0.571$ for partial sky analysis with Mask80. In case of Mask10, the correlations between partial sky $E$- and $B$-modes spectra are within $-0.041$ and $0.997$. These significant correlations (in case of each of these three correlation matrices) are effectively reduced in the full sky spectra predicted by our CNN system for these two polarization modes in each of two different partial sky cases. The correlations in the predicted full sky $E$-mode spectra are shown in the left panel of the bottom row of each of these figures, where the correlations are within $-0.041$ and $0.043$ in case of partial sky analysis with Mask80, and the correlations are in between $\pm 0.042$ for the partial sky analysis with Mask10. In the middle panel of the bottom row of each of these figures, we present the correlation matrix corresponding to the predicted full sky $B$-mode spectra. The correlations coresponding to predicted $B$-mode spectra are in between $-0.047$ and $0.05$ for Mask80, and in case of partial sky analysis with Mask10 these correlations are within $-0.041$ and $0.045$. Moreover, in the right panel of the bottom row of each of these figures, we show the correlations between the predicted full sky spectra of these polarization modes, where the correlations appear within $\pm 0.048$ for Mask80, and in case of Mask10 the range of these correlations is from $-0.046$ to $0.043$. We note that our CNN system provide effectively unbiased predictions of full sky spectra corresponding to the $E$- and $B$-modes, from the noise contaminated partial sky spectra even produced by applying large galactic mask, recovering the $E$-to-$B$ leakage for the multipole range $2 \leq \ell \leq 384$. We also note that the similar correlations in the predicted spectra of these polarization modes for both masks indicates the reliable and well training of our CNN system in case of each of these two different masks. Moreover, the correlations are similar in both cases since our CNN system has trained seperately for both cases which reconstructs the pure $E$- and $B$-modes full sky spectra accurately in both cases even without need of a new complex network for Mask10.
\begin{figure}[h!]
\centering
\includegraphics[scale=0.33]{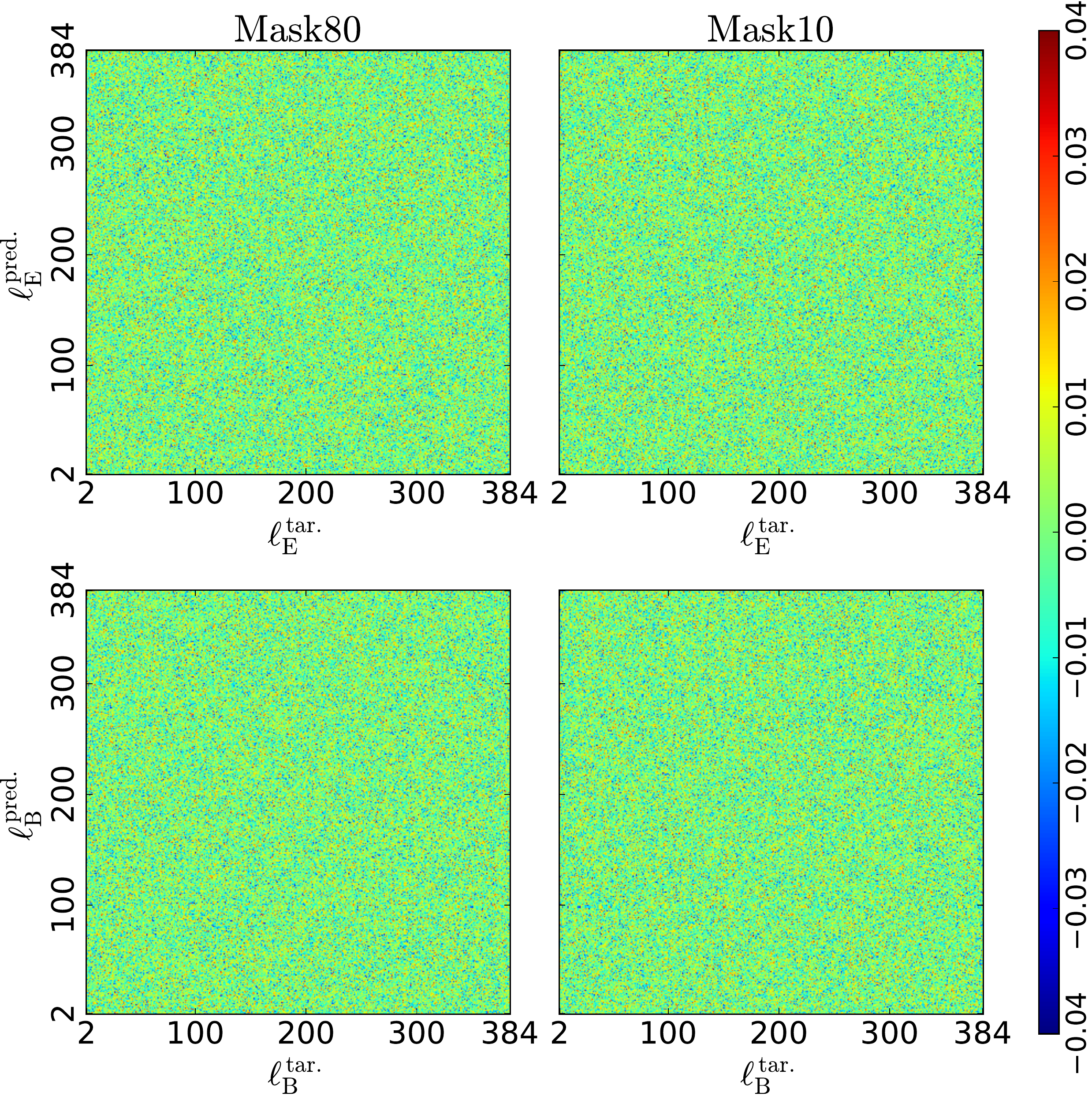}
\caption{Top left and bottom left panels show the correlations between target and predicted spectra for $E$- and $B$-modes respectively for Mask80. Similarly, top right and bottom right panels show the same correlations for Mask10. In each case, the correlations dominantly lie within $\pm 0.04$.}
\label{corr_tar_pred}
\end{figure}

Using equation~\ref{corr_form}, we also calculate the correlations between target and predicted full sky spectra for $E$- and $B$-modes polarizations for each of Mask80 and Mask10 cases. In the top left panel of Figure~\ref{corr_tar_pred}, we show the correlations between the target and predicted $E$-mode spectra for Mask80 case. In the bottom left panel of this figure, we presents the correlations between the target and predicted $B$-mode spectra for the case of Mask80. Similarly, in the top and bottom right panels of this figure, we show these same correlations for $E$- and $B$-modes respectively for Mask10 case. We note that the predicted full sky spectra are largely uncorrelated (i.e., correlations dominantly lie within $\pm 0.04$) with the target spectra for each of these polarization modes for each of these two mask cases. The uncorrelation takes place since we add a random realization component (using predicted aleatoric uncertainty corresponding to the predicted output) during prediction process to preserve the cosmic variance (e.g., see section~\ref{cnn_sec}). The predictions, however, are unbiased and have the same standard deviation as the population of the target spectra, and hence predictions are statistically equivalent to the target spectra.
\section{Discussions and Conclusions}
\label{discussions}
A challenging problem in the contemporary cosmology is accurate detection of primordial $B$-mode polarization. Moreover, the full sky analysis of the observational CMB polarization data can not provide reliable results, since the observed CMB polarized signal may potentially remain contaminated by the residual foregrounds even after a foreground removal method has been performed on the observed maps. Removing the finite contaminated region (i.e., galactic region) of the full sky CMB polarization maps, the partial sky analysis can render approximate information about the early universe. In case of the partial sky CMB polarization analysis, the leakage between the partial sky $E$- and $B$-modes power spectra is a major problem. For removing this leakage problem, the existing classical approaches vary depending upon their limitations and complexities. The work of our current article demonstrates that the ML approach can effectively be a alternative method to extract the full sky information from the partial sky polarization spectra containing `ambiguous' modes and instrumental noises. We note that the  ML system used in this work performs complex algebraic tasks inside the neural network architecture during the process of removing any leakage between $E$- and $B$- modes spectra.

The advantages of our ML approach using the CNN system are as follows. For the particular theoretical spectra of $E$- and $B$-modes, our CNN system can reliably learn the mapping from the noise contaminated partial sky realization spectra to the corresponding noise bias free full sky spectra for each of these polarization modes simultaneously. We have tested our CNN system using two binary masks (i.e., Mask80 and Mask10) seperately. Smoothing of the binary mask is not necessary in our CNN analysis. In case of each of these two different masks, we can apply the binary mask (without any modification) in the noise contaminated full sky $Q$ and $U$ polarization maps for creating the corresponding partial sky $Q$ and $U$ maps. After generating the partial sky $E$- and $B$-modes spectra from these partial sky polarization maps, we train our CNN system using these partial sky spectra labelled by their corresponding full sky spectra (which do not contain detector noises) for a particular mask. We use HS loss function in the training process of our CNN architecture to calculate the aleatoric uncertainties in the CNN system. We also utilize the model averaging ensemble method in the training process to diminish the epistemic uncertainties of the CNN system. The entire training process take only 7 hours in the free GPU platform of \texttt{Google Colab}\footnote{\url{https://colab.research.google.com/}}. This training time is computationally manageable, which is need only once. After the completion of the training process of the CNN system, we can use it multiple time to predict the full sky $E$- and $B$-modes spectra from the unknown partial sky spectra of these modes corresponding to those theoretical spectra for a specific mask. Time taking for the predictions of $10^4$ number of test samples is around two and half minutes in the free GPU platform of \texttt{Google Colab}. These predictions of our CNN system preserve the entire statistical properties (i.e., cosmic variance) of the full sky spectra excellently for both masks. Moreover, the interesting advantage even for Mask10 is that the recovered spectra agree excellently with the actual ground-truths without any need of binning. We note that the predicted spectrum and target spectrum (for the chosen theoretical model) are statistically equivalent to each other in the sense that they have consistent mean and standard deviations. However, the target and the predicted spectra for a given realization are not exactly the same and possess correlation of magnitude less than 5\%. In a future article, we will generalize our method to predict spectra with maximum possible correlations with the targets.

Strong residual foreground contaminations, specifically over very large sky fractions, can be alleviated if one uses a mask that retains only very clean regions of the sky, where foregrounds are less severely contaminating than the weak $B$-mode signal. However, in this case one bears the cost of incurring large sample variance error in the non-ML based methods. These sample variances are more dominant at large angular scales where pristine primordial $B$-mode signal is expected to be observed. Our current work shows that even in presence of realistic sky masks the sample variance issue can be avoided and hence large angular scale primordial $B$-mode signal can be measured reliably (assuming off course foregrounds can be removed reliably from the chosen sky areas). This likely to be another major advantage of using ML method to unmix $E$- and $B$-mode signal from the partial sky.

The working principle of the ML approach is that the ML system learns the relation between the input and output distributions. Our trained CNN system can provide reliable predictions of the full sky spectra for the unknown partial sky realizations. Moreover, we use our CNN system to reconstruct the pure full sky polarization signal from the partial sky data for Mask80 and Mask10 separately. These predictions provide the excellent agreement with the theoretical spectra as well as their cosmic variances for each of these polarization modes for both masks. Moreover, the predictions of our CNN system reliably unbiased, since the significance ratios (section~\ref{sign_sec}) of these predictions are within approximately $3\sigma$ for the multipole range $2 \leq \ell \leq 384$ for both cases. Our CNN system effectively remove the correlations existing in the partial sky spectra of $E$- and $B$-modes polarizations (see section~\ref{corr_sec} for detailed discussions). We note that the CNN model trained in this work uses a given theoretical model for the CMB $E$- and $B$-modes spectra. In reality, the underlying theoretical spectra are not known a priori. In a future article, we will generalize our work to include predictions when the underlying theoretical spectra are unknown. Our current work takes the first major step towards predicting unmixed $E$- and $B$-modes spectra from the partial sky using CNN.

\appendix

\setcounter{section}{0}
\renewcommand\thesection{Appendix A}
\section{: Mixing kernel}
\label{append_kernel}
Using equation~\ref{P_imp_sh} in equations~\ref{E_mode_part} and~\ref{B_mode_part}, the partial sky harmonic coefficients of $E$- and $B$-modes are given by
\begin{eqnarray}
\tilde{a}_{E,\ell m} &=& -\frac{1}{2}\int \sum_{\ell'm'}\Bigl[a^{\rm{imp}}_{2,\ell'm'}Y_{2,\ell'm'}(\hat{n})W(\hat{n})Y^{*}_{2,\ell m}(\hat{n}) + \nonumber \\ 
&& \hspace{25pt} a^{\rm{imp}}_{-2,\ell'm'}Y_{-2,\ell'm'}(\hat{n})W(\hat{n})Y^{*}_{-2,\ell m}(\hat{n})\Bigr]d\Omega \ , \nonumber \\ \label{A1}\\
\tilde{a}_{B,\ell m} &=& \frac{i}{2}\int \sum_{\ell'm'}\Bigl[a^{\rm{imp}}_{2,\ell'm'}Y_{2,\ell'm'}(\hat{n})W(\hat{n})Y^{*}_{2,\ell m}(\hat{n}) - \nonumber \\ 
&& \hspace{25pt} a^{\rm{imp}}_{-2,\ell'm'}Y_{-2,\ell'm'}(\hat{n})W(\hat{n})Y^{*}_{-2,\ell m}(\hat{n})\Bigr]d\Omega \ . \nonumber \\ \label{A2} 
\end{eqnarray}
Using equation~\ref{window} in equations~\ref{A1} and~\ref{A2}, the partial sky harmonic coefficients of these polarization modes can be expressed as
\begin{eqnarray}
\tilde{a}_{E,\ell m} &=& -\frac{1}{2}\sum_{\ell'm'}\Bigl[a^{\rm{imp}}_{2,\ell'm'}W^{(+2)}_{\ell m\ell'm'}+a^{\rm{imp}}_{-2,\ell'm'}W^{(-2)}_{\ell m\ell'm'}\Bigr] \ , \nonumber \\ \label{A3}
\end{eqnarray}
\begin{eqnarray}
\tilde{a}_{B,\ell m} &=& \frac{i}{2}\sum_{\ell'm'}\Bigl[a^{\rm{imp}}_{2,\ell'm'}W^{(+2)}_{\ell m\ell'm'}-a^{\rm{imp}}_{-2,\ell'm'}W^{(-2)}_{\ell m\ell'm'}\Bigr] \ . \nonumber \\ \label{A4}
\end{eqnarray}
We can simplify these equations~\ref{A3} and~\ref{A4} as
\begin{eqnarray}
\tilde{a}_{E,\ell m} &=& \sum_{\ell'm'}\Biggl[\Biggl\{\frac{1}{2}\Bigl(W^{(+2)}_{\ell m\ell'm'}+W^{(-2)}_{\ell m\ell'm'}\Bigr)\Biggr\} \times \nonumber \\
&& \hspace{20pt} \Biggl\{-\frac{1}{2}\Bigl(a^{\rm{imp}}_{2,\ell'm'}+a^{\rm{imp}}_{-2,\ell'm'}\Bigr)\Biggr\}\Biggr]+ \nonumber \\ 
&& \sum_{\ell'm'}i\Biggl[\Biggl\{\frac{1}{2}\Bigl(W^{(+2)}_{\ell m\ell'm'}-W^{(-2)}_{\ell m\ell'm'}\Bigr)\Biggr\} \times \nonumber \\ 
&& \hspace{25pt} \Biggl\{\frac{i}{2}\Bigl(a^{\rm{imp}}_{2,\ell'm'}-a^{\rm{imp}}_{-2,\ell'm'}\Bigr)\Biggr\}\Biggr] \nonumber \\[10pt] &=& \sum_{\ell'm'}\Bigl[K^{(+)}_{\ell m\ell'm'}a^{\rm{imp}}_{E,\ell'm'}+iK^{(-)}_{\ell m\ell'm'}a^{\rm{imp}}_{B,\ell'm'}\Bigr] \ , \label{A5}
\end{eqnarray}
\begin{eqnarray}
\tilde{a}_{B,\ell m} &=& \sum_{\ell'm'}\Biggl[\Biggl\{\frac{1}{2}\Bigl(W^{(+2)}_{\ell m\ell'm'}+W^{(-2)}_{\ell m\ell'm'}\Bigr)\Biggr\} \times \nonumber \\
&& \hspace{20pt} \Biggl\{\frac{i}{2}\Bigl(a^{\rm{imp}}_{2,\ell'm'}-a^{\rm{imp}}_{-2,\ell'm'}\Bigr)\Biggr\}\Biggr]- \nonumber \\ & & \sum_{\ell'm'}i\Biggl[\Biggl\{\frac{1}{2}\Bigl(W^{(+2)}_{\ell m\ell'm'}-W^{(-2)}_{\ell m\ell'm'}\Bigr)\Biggr\} \times \nonumber \\
&& \hspace{25pt} \Biggl\{-\frac{1}{2}\Bigl(a^{\rm{imp}}_{2,\ell'm'}+a^{\rm{imp}}_{-2,\ell'm'}\Bigr)\Biggr\}\Biggr] \nonumber \\[10pt] &=& \sum_{\ell'm'}\Bigl[K^{(+)}_{\ell m\ell'm'}a^{\rm{imp}}_{B,\ell'm'}-iK^{(-)}_{\ell m\ell'm'}a^{\rm{imp}}_{E,\ell'm'}\Bigr] \ , \label{A6}
\end{eqnarray}
where the mixing kernels are defined as 
\begin{eqnarray}
K^{(\pm)}_{\ell m\ell' m'} &=& \frac{1}{2}\Bigl[W^{(+2)}_{\ell m\ell' m'}\pm W^{(-2)}_{\ell m\ell' m'}\Bigr] \ . \label{A7}
\end{eqnarray}

\appendix

\setcounter{section}{1}
\renewcommand\thesection{Appendix B}
\section{: Mixing matrix}
\label{append_matrix}
Ensemble average of the realizations of the power spectrum is given by
\begin{eqnarray}
\bigl<C_{l}\bigr> &=& \frac{1}{2\ell +1}\sum_{m}\bigl<a_{\ell m}a^{*}_{\ell m}\bigr> \ .\label{B1}
\end{eqnarray}
Moreover, since the ensemble average of the entire realizations of the power spectrum deliver the theoretical power spectrum, the ensemble average of the multiplication between harmonic coefficient and its complex conjugate is given by
\begin{eqnarray}
\bigl<a_{\ell m}a^{*}_{\ell'm'}\bigr> &=& \bigl<C_{l}\bigr>\delta_{\ell\ell'}\delta_{mm'} \ ,\label{B2}
\end{eqnarray}
where $\delta$ defines the Kronecker delta function.

The complex conjugate expressions of equations~\ref{A5} and~\ref{A6} are given by
\begin{eqnarray}
\tilde{a}^{*}_{E,\ell m} &=& \sum_{\ell'm'}\Bigl[K^{(+)*}_{\ell m\ell'm'}a^{\rm{imp*}}_{E,\ell'm'}-iK^{(-)*}_{\ell m\ell'm'}a^{\rm{imp*}}_{B,\ell'm'}\Bigr] \ , \label{B3} \\
\tilde{a}^{*}_{B,\ell m} &=& \sum_{\ell'm'}\Bigl[K^{(+)*}_{\ell m\ell'm'}a^{\rm{imp*}}_{B,\ell'm'}+iK^{(-)*}_{\ell m\ell'm'}a^{\rm{imp*}}_{E,\ell'm'}\Bigr] \ . \label{B4}
\end{eqnarray}
The $E$-mode polarization has the opposite parity to the $B$-mode polarization. This provides
\begin{eqnarray}
\bigl<a_{E,\ell m}a^{*}_{B,\ell'm'}\bigr> = \bigl<a_{B,\ell m}a^{*}_{E,\ell'm'}\bigr> = 0 \ . \label{B5}
\end{eqnarray}
Therefore, using equations~\ref{noiseEB_cross},~\ref{E_mode_imp},~\ref{B_mode_imp} and~\ref{B5} we get
\begin{eqnarray}
\bigl<a^{\rm{imp}}_{E,\ell m}a^{\rm{imp*}}_{B,\ell'm'}\bigr> = \bigl<a^{\rm{imp}}_{B,\ell m}a^{\rm{imp*}}_{E,\ell'm'}\bigr> = 0 \ . \label{B6}
\end{eqnarray}
Taking the multiplication of the equation~\ref{A5} with the equation~\ref{B3} as well as the multiplication of the equation~\ref{A6} with the equation~\ref{B4} and also using the equation~\ref{B6}, we can express the ensemble average of these multiplications between the partial sky harmonic mode and its complex conjugate. This expression corresponding to $E$-mode polarization is written as
\begin{eqnarray}
\bigl<\tilde{a}_{E,\ell m}\tilde{a}^{*}_{E,\ell m}\bigr> &=& \sum_{\ell'm'}\sum_{\ell''m''}\Bigl[\Bigl\{K^{(+)}_{\ell m\ell'm'}K^{(+)*}_{\ell m\ell''m''} \times \nonumber \\
&& \hspace{40pt} \bigl<a^{\rm{imp}}_{E,\ell'm'}a^{\rm{imp*}}_{E,\ell''m''}\bigr>\Bigr\}+ \nonumber \\ 
&& \hspace{40pt} \Bigl\{K^{(-)}_{\ell m\ell'm'}K^{(-)*}_{\ell m\ell''m''} \times \nonumber \\
&& \hspace{40pt} \bigl<a^{\rm{imp}}_{B,\ell'm'}a^{\rm{imp*}}_{B,\ell''m''}\bigr>\Bigr\}\Bigr] \ . \label{B7}
\end{eqnarray}
Using equations~\ref{noiseE_auto},~\ref{noiseB_auto},~\ref{E_mode_imp},~\ref{B_mode_imp} and~\ref{B2}, the equation~\ref{B7} can be simplified as
\begin{eqnarray}
\bigl<\tilde{a}_{E,\ell m}\tilde{a}^{*}_{E,\ell m}\bigr> &=& \sum_{\ell'm'}\sum_{\ell''m''}\Biggl[\Biggr\{K^{(+)}_{\ell m\ell'm'}K^{(+)*}_{\ell m\ell''m''} \times \nonumber \\
&& \hspace{10pt} \left(\bigl<C^{EE}_{\ell'}\bigr>+\frac{\bigl<N_{\ell'}^{EE}\bigr>}{\mathcal{B}_{E}^{2}(\ell')}\right)\delta_{\ell'\ell''}\delta_{m'm''}\Biggr\}+ \nonumber \\ 
&& \hspace{10pt} \Biggl\{K^{(-)}_{\ell m\ell'm'}K^{(-)*}_{\ell m\ell''m''} \times \nonumber \\
&& \hspace{10pt} \left(\bigl<C^{BB}_{\ell'}\bigr>+\frac{\bigl<N_{\ell'}^{BB}\bigr>}{\mathcal{B}_{B}^{2}(\ell')}\right)\delta_{\ell'\ell''}\delta_{m'm''}\Biggr\}\Biggr] \nonumber \\ &=& \sum_{\ell'm'}\Biggl[|K^{(+)}_{\ell m\ell'm'}|^{2}\left(\bigl<C^{EE}_{\ell'}\bigr>+\frac{\bigl<N_{\ell'}^{EE}\bigr>}{\mathcal{B}_{E}^{2}(\ell')}\right) + \nonumber \\ 
&& \hspace{10pt} |K^{(-)}_{\ell m\ell'm'}|^{2}\left(\bigl<C^{BB}_{\ell'}\bigr>+\frac{\bigl<N_{\ell'}^{BB}\bigr>}{\mathcal{B}_{B}^{2}(\ell')}\right)\Biggr] \ . \label{B8}
\end{eqnarray}
Multiplying the both sides of the equation~\ref{B8} by $\frac{1}{2\ell +1}$ after taking the sum over $m$ index and then using equation~\ref{B1}, the more simplified form of equation~\ref{B8} is expressed as
\begin{eqnarray}
\bigl<\tilde{C}^{EE}_{\ell}\bigr> &=& \sum_{\ell'}\sum_{mm'}\Biggl[\frac{|K^{(+)}_{\ell m\ell'm'}|^{2}}{2\ell +1}\left(\bigl<C^{EE}_{\ell'}\bigr>+\frac{\bigl<N_{\ell'}^{EE}\bigr>}{\mathcal{B}_{E}^{2}(\ell')}\right) + \nonumber \\
&& \hspace{40pt} \frac{|K^{(-)}_{\ell m\ell'm'}|^{2}}{2\ell +1}\left(\bigl<C^{BB}_{\ell'}\bigr>+\frac{\bigl<N_{\ell'}^{BB}\bigr>}{\mathcal{B}_{B}^{2}(\ell')}\right)\Biggr] \nonumber \\ 
&=& \sum_{\ell'}\Biggl[M^{(+)}_{\ell\ell'}\left(\bigl<C^{EE}_{\ell'}\bigr>+\frac{\bigl<N_{\ell'}^{EE}\bigr>}{\mathcal{B}_{E}^{2}(\ell')}\right) + \nonumber \\
&& \hspace{20pt} M^{(-)}_{\ell\ell'}\left(\bigl<C^{BB}_{\ell'}\bigr>+\frac{\bigl<N_{\ell'}^{BB}\bigr>}{\mathcal{B}_{B}^{2}(\ell')}\right)\Biggr] \ , \label{B9}
\end{eqnarray}
where
\begin{eqnarray}
\bigl<\tilde{C}^{EE}_{\ell}\bigr> &=& \frac{1}{2\ell +1}\sum_{m}\bigl<\tilde{a}_{E,\ell m}\tilde{a}^{*}_{E,\ell m}\bigr>. \label{B10}
\end{eqnarray} 
In equation~\ref{B9}, the mixing matrices are defined as
\begin{eqnarray}
M_{\ell \ell'}^{(\pm)} &=& \sum_{mm'}\frac{1}{2\ell +1}|K^{(\pm)}_{\ell m\ell' m'}|^2 \ . \label{B11}
\end{eqnarray}
Equation~\ref{B9} denotes the relation between the full sky and the partial sky $E$-mode power spectra. Following the similar procedure, the partial sky power spectrum corresponding to the $B$-mode polarization, in terms of full sky spectrum, can be expressed by
\begin{eqnarray}
\bigl<\tilde{C}^{BB}_{\ell}\bigr> &=& \sum_{\ell'}\Biggl[M^{(+)}_{\ell\ell'}\left(\bigl<C^{BB}_{\ell'}\bigr>+\frac{\bigl<N_{\ell'}^{BB}\bigr>}{\mathcal{B}_{B}^{2}(\ell')}\right)+ \nonumber \\ 
&& \hspace{20pt} M^{(-)}_{\ell\ell'}\left(\bigl<C^{EE}_{\ell'}\bigr>+\frac{\bigl<N_{\ell'}^{EE}\bigr>}{\mathcal{B}_{E}^{2}(\ell')}\right)\Biggr] \ . \label{B12}
\end{eqnarray}

\section*{Acknowledgement}
\label{acknowledgement}
We acknowledge the use of the openly available packages CAMB\footnote{\url{https://camb.readthedocs.io/en/latest/}}, HEALPix\footnote{\url{https://healpix.sourceforge.io/}}, \texttt{healpy}\footnote{\url{https://github.com/healpy/healpy}}, TensorFlow\footnote{\url{https://www.tensorflow.org/}} and \texttt{Google Colab}\footnote{\url{https://colab.research.google.com/}}. We thank to Ujjal Purkayastha, Sarvesh Kumar Yadav and Albin Joseph for constructive discussions related to this work.


\begin{thebibliography}{}

\makeatletter
\relax
\def\mn@urlcharsother{\let\do\@makeother \do\$\do\&\do\#\do\^\do\_\do\%\do\~}
\def\mn@doi{\begingroup\mn@urlcharsother \@ifnextchar [ {\mn@doi@}
  {\mn@doi@[]}}
\def\mn@doi@[#1]#2{\def\@tempa{#1}\ifx\@tempa\@empty \href
  {http://dx.doi.org/#2} {doi:#2}\else \href {http://dx.doi.org/#2} {#1}\fi
  \endgroup}
\def\mn@eprint#1#2{\mn@eprint@#1:#2::\@nil}
\def\mn@eprint@arXiv#1{\href {http://arxiv.org/abs/#1} {{\tt arXiv:#1}}}
\def\mn@eprint@dblp#1{\href {http://dblp.uni-trier.de/rec/bibtex/#1.xml}
  {dblp:#1}}
\def\mn@eprint@#1:#2:#3:#4\@nil{\def\@tempa {#1}\def\@tempb {#2}\def\@tempc
  {#3}\ifx \@tempc \@empty \let \@tempc \@tempb \let \@tempb \@tempa \fi \ifx
  \@tempb \@empty \def\@tempb {arXiv}\fi \@ifundefined
  {mn@eprint@\@tempb}{\@tempb:\@tempc}{\expandafter \expandafter \csname
  mn@eprint@\@tempb\endcsname \expandafter{\@tempc}}}

\bibitem[\protect\citeauthoryear{{Abadi} et al.}{2015}]{Abadi_2015}
Abadi M., et al., \emph{TensorFlow: Large-Scale Machine Learning on Heterogeneous Systems}, 2015,
\href{http://download.tensorflow.org/paper/whitepaper2015.pdf}{download.tensorflow.org/paper/whitepaper2015.pdf}

\bibitem[\protect\citeauthoryear{{Planck} {Collaboration} {I}}{2020}]{PlanckI_2020}
Aghanim, N. et al., \emph{Planck 2018 results: I. Overview and the cosmological legacy of Planck}, 2020, \mn@doi [A\&A]
{10.1051/0004-6361/201833880}, \href
{https://doi.org/10.1051/0004-6361/201833880}{641, A1}

\bibitem[\protect\citeauthoryear{{Planck} {Collaboration} {V}}{2020}]{PlanckV_2020}
Aghanim, N. et al., \emph{Planck 2018 results: V. CMB power spectra and likelihoods}, 2020, \mn@doi [A\&A]
{10.1051/0004-6361/201936386}, \href
{https://doi.org/10.1051/0004-6361/201936386}{641, A5}

\bibitem[\protect\citeauthoryear{{Planck} {Collaboration} {VI}}{2020}]{PlanckVI_2020}
Aghanim, N. et al., \emph{Planck 2018 results: VI. Cosmological parameters}, 2020, \mn@doi [A\&A]
{10.1051/0004-6361/201833910}, \href
{https://doi.org/10.1051/0004-6361/201833910}{641, A6}

\bibitem[\protect\citeauthoryear{{Aiola} et al.}{2020}]{Aiola_2020}
Aiola, S. et al., \emph{The Atacama Cosmology Telescope: DR4 maps and cosmological parameters}, 2020, \mn@doi [JCAP]
{10.1088/1475-7516/2020/12/047}, \href
{https://doi.org/10.1088/1475-7516/2020/12/047}{2020, 047}

\bibitem[\protect\citeauthoryear{{Planck} {Collaboration} {IV}}{2020}]{PlanckIV_2020}
Akrami, Y. et al., \emph{Planck 2018 results: IV. Diffuse component separation}, 2020, \mn@doi [A\&A]
{10.1051/0004-6361/201833881}, \href
{https://doi.org/10.1051/0004-6361/201833881}{641, A4}

\bibitem[\protect\citeauthoryear{{Alonso} et al.}{2019}]{Alonso_2019}
Alonso, D., Sanchez, J., Slosar, A. and LSST Dark Energy Science Collaboration, \emph{A unified pseudo-$C_{\ell}$ framework}, 2019, \mn@doi [MNRAS]
{10.1093/mnras/stz093}, \href
{https://doi.org/10.1093/mnras/stz093}{484, 4127}

\bibitem[\protect\citeauthoryear{{Baccigalupi} et al.}{2000}]{Baccigalupi_2000}
Baccigalupi C., Bedini L., Burigana C., de Zotti G., Farusi A., Maino D., Maris M., Perrotta F., Salerno E., Toffolatti L. and Tonazzini A., \emph{Neural networks and the separation of cosmic microwave background and astrophysical signals in sky maps}, 2000, \mn@doi [MNRAS]
{10.1046/j.1365-8711.2000.03751.x}, \href
{https://doi.org/10.1046/j.1365-8711.2000.03751.x}{318, 769}

\bibitem[\protect\citeauthoryear{{Bennett} et al.}{1996}]{Bennett_1996}
Bennett, C. L. et al., \emph{Four-Year COBE DMR Cosmic Microwave Background Observations: Maps and Basic Results}, 1996, \mn@doi [ApJ]
{10.1086/310075}, \href
{https://doi.org/10.1086/310075}{464, L1}

\bibitem[\protect\citeauthoryear{{Bennett} et al.}{2013}]{Bennett_2013}
Bennett, C. L. et al., \emph{NINE-YEAR WILKINSON MICROWAVE ANISOTROPY PROBE (WMAP) OBSERVATIONS: FINAL MAPS AND RESULTS}, 2013, \mn@doi [ApJS]
{10.1088/0067-0049/208/2/20}, \href
{https://doi.org/10.1088/0067-0049/208/2/20}{208, 20}

\bibitem[\protect\citeauthoryear{{Benson} et al.}{2014}]{Benson_2014}
Benson, B. et al., \emph{SPT-3G: a next-generation cosmic microwave background polarization experiment on the South Pole telescope}, 2014, \mn@doi [SPIE Proceedings]
{10.1117/12.2057305}, \href
{https://lens.org/003-989-902-450-892}{9153, 552}

\bibitem[\protect\citeauthoryear{{Brown} et al.}{2009}]{Brown_2009}
Brown, M. L. et al., \emph{IMPROVED MEASUREMENTS OF THE TEMPERATURE AND POLARIZATION OF THE COSMIC MICROWAVE BACKGROUND FROM QUaD}, 2009, \mn@doi [ApJ]
{10.1088/0004-637x/705/1/978}, \href
{https://doi.org/10.1088/0004-637x/705/1/978}{705, 978}

\bibitem[\protect\citeauthoryear{{Bunn} et al.}{2003}]{Bunn_2003}
Bunn, E., F., Zaldarriaga, M., Tegmark, M. and de Oliveira-Costa, A., \emph{E/B decomposition of finite pixelized CMB maps}, 2003, \mn@doi [Phys. Rev. D]
{10.1103/PhysRevD.67.023501}, \href
{https://link.aps.org/doi/10.1103/PhysRevD.67.023501}{67, 023501}

\bibitem[\protect\citeauthoryear{{Bunn} \& {Wandelt}}{2017}]{Bunn_2017}
Bunn, E., F. and Wandelt, B., \emph{Pure E and B polarization maps via Wiener filtering}, 2017, \mn@doi [Phys. Rev. D]
{10.1103/PhysRevD.96.043523}, \href
{https://link.aps.org/doi/10.1103/PhysRevD.96.043523}{96, 043523}

\bibitem[\protect\citeauthoryear{{Chanda} \& {Saha}}{2021}]{Chanda_2021}
Chanda P. and Saha R., \emph{An unbiased estimator of the full-sky CMB angular power spectrum at large scales using neural networks}, 2021, \mn@doi [MNRAS]
{10.1093/mnras/stab2753}, \href
{https://doi.org/10.1093/mnras/stab2753}{508, 4600}

\bibitem[\protect\citeauthoryear{{Chiang} et al.}{2010}]{Chiang_2010}
Chiang, H. C. et al., \emph{MEASUREMENT OF COSMIC MICROWAVE BACKGROUND POLARIZATION POWER SPECTRA FROM TWO YEARS OF BICEP DATA}, 2010, \mn@doi [ApJ]
{10.1088/0004-637x/711/2/1123}, \href
{https://doi.org/10.1088/0004-637x/711/2/1123}{711, 1123}

\bibitem[\protect\citeauthoryear{{Choi} et al.}{2020}]{Choi_2020}
Choi, S. K. et al., \emph{The Atacama Cosmology Telescope: a measurement of the Cosmic Microwave Background power spectra at 98 and 150 GHz}, 2020, \mn@doi [JCAP]
{10.1088/1475-7516/2020/12/045}, \href
{https://doi.org/10.1088/1475-7516/2020/12/045}{2020, 045}

\bibitem[\protect\citeauthoryear{{Delabrouille} et al.}{2018}]{Delabrouille_2018}
Delabrouille, J. et al., \emph{Exploring cosmic origins with CORE: Survey requirements and mission design}, 2018, \mn@doi [JCAP]
{10.1088/1475-7516/2018/04/014}, \href
{https://doi.org/10.1088/1475-7516/2018/04/014}{2018, 014}

\bibitem[\protect\citeauthoryear{{Dialektopoulos} et al.}{2021}]{Dialektopoulos_2021}
Dialektopoulos K., Said J. L., Mifsud J., Sultana J. and Adami K. Z., \emph{Neural Network Reconstruction of Late-Time Cosmology and Null Tests}, 2021,
\href{https://arxiv.org/abs/2111.11462}{arxiv.org/abs/2111.11462}

\bibitem[\protect\citeauthoryear{{Dodelson} \& {Schmidt}}{2021}]{Dodelson_2021}
Dodelson S. and Schmidt F., \emph{MODERN COSMOLOGY}, 2021.

\bibitem[\protect\citeauthoryear{{Dutcher} et al.}{2021}]{Dutcher_2021}
Dutcher, D. et al., \emph{Measurements of the E-mode polarization and temperature-E-mode correlation of the CMB from SPT-3G 2018 data}, 2021, \mn@doi [Phys. Rev. D]
{10.1103/PhysRevD.104.022003}, \href
{https://link.aps.org/doi/10.1103/PhysRevD.104.022003}{104, 022003}

\bibitem[\protect\citeauthoryear{{Escamilla-Rivera} et al.}{2020}]{Escamilla_2020}
Escamilla-Rivera C., Quintero M. A. C. and Capozziello S., \emph{A deep learning approach to cosmological dark energy models}, 2020, \mn@doi [JCAP]
{10.1088/1475-7516/2020/03/008}, \href
{https://doi.org/10.1088/1475-7516/2020/03/008}{2020, 008}

\bibitem[\protect\citeauthoryear{{Fert\'e} et al.}{2013}]{Ferte_2013}
Fert\'e, A., Grain, J., Tristram, M. and Stompor, R., \emph{Efficiency of pseudospectrum methods for estimation of the cosmic microwave background B-mode power spectrum}, 2013, \mn@doi [Phys. Rev. D]
{10.1103/PhysRevD.88.023524}, \href
{https://link.aps.org/doi/10.1103/PhysRevD.88.023524}{88, 023524}

\bibitem[\protect\citeauthoryear{{Fixsen} et al.}{1994}]{Fixsen_1994}
Fixsen D. J. et al., \emph{Cosmic Microwave Background Dipole Spectrum Measured by the COBE FIRAS Instrument}, 1994, \mn@doi [ApJ]
{10.1086/173575}, \href
{https://ui.adsabs.harvard.edu/abs/1994ApJ...420..445F}{420, 445}

\bibitem[\protect\citeauthoryear{{Fixsen} et al.}{1996}]{Fixsen_1996}
Fixsen D. J., Cheng E. S., Gales J. M., Mather J. C., Shafer R. A. and Wright E. L., \emph{The Cosmic Microwave Background Spectrum from the Full COBE FIRAS Data Set}, 1996, \mn@doi [ApJ]
{10.1086/178173}, \href
{https://doi.org/10.1086/178173}{473, 576}

\bibitem[\protect\citeauthoryear{{Flöss} \& {Meerburg}}{2023}]{Floss_2023}
Flöss T. and Meerburg P. D., \emph{Improving constraints on primordial non-Gaussianity using neural network based reconstruction}, 2023, 
\href{https://arxiv.org/abs/2305.07018}{arXiv:2305.07018}

\bibitem[\protect\citeauthoryear{{G{\'{o}}mez-Vargas}, {Esquivel} et al.}{2021}]{Gomez1_2021}
G{\'{o}}mez-Vargas I., Esquivel R. M., Garc{\'{\i}}a-Salcedo R. and V{\'{a}}zquez J. A., \emph{Neural network within a Bayesian inference framework}, 2021, \mn@doi [Journal of Physics: Conference Series]
{10.1088/1742-6596/1723/1/012022}, \href
{https://doi.org/10.1088/1742-6596/1723/1/012022}{1723, 012022}

\bibitem[\protect\citeauthoryear{{G{\'{o}}mez-Vargas}, {Vázquez} et al.}{2021}]{Gomez2_2021}
G{\'{o}}mez-Vargas I., Vázquez J. A., Esquivel R. M. and García-Salcedo R., \emph{Neural network reconstructions for the Hubble parameter, growth rate and distance modulus}, 2021, 
\href{https://arxiv.org/abs/2104.00595}{arxiv.org/abs/2104.00595}

\bibitem[\protect\citeauthoryear{{Gorski} et al.}{2005}]{Gorski_2005}
Gorski M. K., Hivon E., Banday J. A., Wandelt D. B., Hansen K. F., Reinecke M. and Bartelmann M., \emph{HEALPix: A Framework for High-Resolution Discretization and Fast Analysis of Data Distributed on the Sphere}, 2005, \mn@doi [ApJ]
{10.1086/427976}, \href
{https://doi.org/10.1086/427976}{622, 759}

\bibitem[\protect\citeauthoryear{{Graff} et al.}{2012}]{Graff_2012}
Graff P., Feroz F., Hobson M. P. and Lasenby A., \emph{BAMBI: blind accelerated multimodal Bayesian inference}, 2012, \mn@doi [MNRAS]
{10.1111/j.1365-2966.2011.20288.x}, \href
{https://doi.org/10.1111/j.1365-2966.2011.20288.x}{421, 169}

\bibitem[\protect\citeauthoryear{{Gu} et al.}{2015}]{Gu_2015}
Gu, J., et al., \emph{Recent Advances in Convolutional Neural Networks}, 2015,
\href{https://arxiv.org/abs/1512.07108}{arxiv.org/abs/1512.07108}

\bibitem[\protect\citeauthoryear{{Guth} \& {Pi}}{1982}]{Guth_1982}
Guth, Alan H. and Pi, So-Young, \emph{Fluctuations in the New Inflationary Universe}, 1982, \mn@doi [Phys. Rev. Lett.]
{10.1103/PhysRevLett.49.1110}, \href
{https://link.aps.org/doi/10.1103/PhysRevLett.49.1110}{49, 1110}

\bibitem[\protect\citeauthoryear{{Hanany} et al.}{2019}]{Hanany_2019}
Hanany, S. et al., \emph{PICO: Probe of Inflation and Cosmic Origins}, 2019, 
\href{https://arxiv.org/pdf/1902.10541.pdf}{arxiv.org/pdf/1902.10541.pdf}

\bibitem[\protect\citeauthoryear{{Hazumi} et al.}{2020}]{Hazumi_2020}
Hazumi, M. et al., \emph{LiteBIRD satellite: JAXA’s new strategic L-class mission for
all-sky surveys of cosmic microwave background polarization}, 2020,
\href{https://arxiv.org/pdf/2101.12449.pdf}{arxiv.org/pdf/2101.12449.pdf}

\bibitem[\protect\citeauthoryear{{Hinshaw} et al.}{2013}]{Hinshaw_2013}
Hinshaw, G. et al., \emph{NINE-YEAR WILKINSON MICROWAVE ANISOTROPY PROBE (WMAP) OBSERVATIONS: COSMOLOGICAL PARAMETER RESULTS}, 2013, \mn@doi [ApJS]
{10.1088/0067-0049/208/2/19}, \href
{https://doi.org/10.1088/0067-0049/208/2/19}{208, 19}

\bibitem[\protect\citeauthoryear{{Hortua} et al.}{2020}]{Hortua_2020}
Hortua H. J., Volpi R., Marinelli D. and Malago L., \emph{Accelerating MCMC algorithms through Bayesian Deep Networks}, 2020,
\href{https://arxiv.org/abs/2011.14276}{arxiv.org/abs/2011.14276}

\bibitem[\protect\citeauthoryear{{Hou} et al.}{2014}]{Hou_2014}
Hou Z. et al., \emph{CONSTRAINTS ON COSMOLOGY FROM THE COSMIC MICROWAVE BACKGROUND POWER SPECTRUM OF THE 2500 deg$^2$ SPT-SZ SURVEY}, 2014, \mn@doi [ApJ]
{10.1088/0004-637x/782/2/74}, \href
{https://doi.org/10.1088/0004-637x/782/2/74}{782, 74}

\bibitem[\protect\citeauthoryear{{Hu} \& {White}}{1997}]{Hu_1997}
Hu, W. and White, M., \emph{CMB anisotropies: Total angular momentum method}, 1997, \mn@doi [Phys. Rev. D]
{10.1103/PhysRevD.56.596}, \href
{https://link.aps.org/doi/10.1103/PhysRevD.56.596}{56, 596}

\bibitem[\protect\citeauthoryear{{Kamionkowski} et al.}{1997}]{Kamionkowski_1997}
Kamionkowski, M., Kosowsky, A. and Stebbins, A., \emph{Statistics of cosmic microwave background polarization}, 1997, \mn@doi [Phys. Rev. D]
{10.1103/PhysRevD.55.7368}, \href
{https://link.aps.org/doi/10.1103/PhysRevD.55.7368}{55, 7368}

\bibitem[\protect\citeauthoryear{{Kamionkowski} \& {Kovetz}}{2016}]{Kamionkowski_2016}
Kamionkowski, M. and Kovetz, E., D., \emph{The Quest for B Modes from Inflationary Gravitational Waves}, 2016, \mn@doi [ARA\&A]
{10.1146/annurev-astro-081915-023433}, \href
{https://doi.org/10.1146/annurev-astro-081915-023433}{54, 227}

\bibitem[\protect\citeauthoryear{{Kendall} {\&} {Gal}}{2017}]{Kendall_2017}
Kendall A., Gal Y., \emph{What Uncertainties Do We Need in Bayesian Deep Learning for Computer Vision?}, 2017,
\href{https://arxiv.org/abs/1703.04977.pdf}{arxiv.org/abs/1703.04977.pdf}

\bibitem[\protect\citeauthoryear{{Khan} \& {Saha}}{2023}]{Khan_2023}
{Khan, M. I.} and {Saha, R.}, \emph{Detection of Dipole Modulation in CMB Temperature Anisotropy Maps from WMAP and Planck using Artificial Intelligence}, 2023, \mn@doi [ApJ]
{10.3847/1538-4357/acbfa9}, \href
{https://dx.doi.org/10.3847/1538-4357/acbfa9}{947, 47}

\bibitem[\protect\citeauthoryear{{Kim} \& {Naselsky}}{2010}]{Kim_2010}
{Kim, J.} and {Naselsky, P.}, \emph{E/B decomposition of CMB polarization pattern of incomplete sky: a pixel space approach}, 2010, \mn@doi [A\&A]
{10.1051/0004-6361/201014739}, \href
{https://doi.org/10.1051/0004-6361/201014739}{519, A104}

\bibitem[\protect\citeauthoryear{{Kingma} {\&} {Ba}}{2014}]{Kingma_2014}
Kingma D. P., Ba J., \emph{Adam: A method for stochastic optimization}, 2014,
\href{https://arxiv.org/abs/1412.6980}{arxiv.org/abs/1412.6980}

\bibitem[\protect\citeauthoryear{{Kogut} et al.}{2016}]{Kogut_2016}
Kogut, A., \emph{The Primordial Inflation Explorer (PIXIE)}, 2016, \mn@doi [Society of Photo-Optical Instrumentation Engineers (SPIE) Conference Series]
{10.1117/12.2231090}, \href
{https://ui.adsabs.harvard.edu/abs/2016SPIE.9904E..0WK}{9904, 99040W}

\bibitem[\protect\citeauthoryear{{Kovac} et al.}{2002}]{Kovac_2002}
Kovac, J., M., \emph{Detection of polarization in the cosmic microwave background using DASI}, 2002, \mn@doi [Nature]
{10.1038/nature01269}, \href
{https://doi.org/10.1038/nature01269}{420, 772}

\bibitem[\protect\citeauthoryear{{Lai} et al.}{2021}]{Lai_2021}
Lai Y., Shi Y., Han Y., Shao Y., Qi M. and Li B., \emph{Exploring Uncertainty in Deep Learning for Construction of Prediction Intervals}, 2021,
\href{https://arxiv.org/abs/2104.12953}{arxiv.org/abs/2104.12953}

\bibitem[\protect\citeauthoryear{{Leitch} et al.}{2002}]{Leitch_2002}
Leitch, E., M., \emph{Measurement of polarization with the Degree Angular Scale Interferometer}, 2002, \mn@doi [Nature]
{10.1038/nature01271}, \href
{https://doi.org/10.1038/nature01271}{420, 763}

\bibitem[\protect\citeauthoryear{{Lewis} et al.}{2001}]{Lewis_2001}
Lewis, A., Challinor, A. and Turok, N., \emph{Analysis of CMB polarization on an incomplete sky}, 2001, \mn@doi [Phys. Rev. D]
{10.1103/PhysRevD.65.023505}, \href
{https://link.aps.org/doi/10.1103/PhysRevD.65.023505}{65, 023505}

\bibitem[\protect\citeauthoryear{{Lewis}}{2003}]{Lewis_2003}
Lewis, A., \emph{Harmonic E/B decomposition for CMB polarization maps}, 2003, \mn@doi [Phys. Rev. D]
{10.1103/PhysRevD.68.083509}, \href
{https://link.aps.org/doi/10.1103/PhysRevD.68.083509}{68, 083509}

\bibitem[\protect\citeauthoryear{{Liu}, {Creswell} \& {Dachlythra}}{2019}]{Liu2_2019}
Liu, H., Creswell, J. and Dachlythra, K., \emph{Blind correction of the EB-leakage in the pixel domain}, 2019, \mn@doi [JCAP]
{10.1088/1475-7516/2019/04/046}, \href
{https://doi.org/10.1088/1475-7516/2019/04/046}{04, 046}

\bibitem[\protect\citeauthoryear{{Liu} et al.}{2019}]{Liu_2019}
Liu, H., Creswell, J., von Hausegger, S. and Naselsky, P., \emph{Methods for pixel domain correction of EB leakage}, 2019, \mn@doi [Phys. Rev. D]
{10.1103/PhysRevD.100.023538}, \href
{https://link.aps.org/doi/10.1103/PhysRevD.100.023538}{100, 023538}

\bibitem[\protect\citeauthoryear{{Mancini} et al.}{2022}]{Mancini_2022}
Mancini A. S., Piras D., Alsing J., Joachimi B. and Hobson M. P., \emph{CosmoPower: emulating cosmological power spectra for accelerated Bayesian inference from next-generation surveys}, 2022, \mn@doi [MNRAS]
{10.1093/mnras/stac064}, \href
{https://doi.org/10.1093/mnras/stac064}{511, 1771}

\bibitem[\protect\citeauthoryear{{Mather} et al.}{1999}]{Mather_1999}
Mather J. C., Fixsen D. J., Shafer R. A., Mosier C. and Wilkinson D. T., \emph{Calibrator Design for the COBE Far Infrared Absolute Spectrophotometer (FIRAS)}, 1999, \mn@doi [ApJ]
{10.1086/306805}, \href
{https://doi.org/10.1086/306805}{512, 511}

\bibitem[\protect\citeauthoryear{{Moss}}{2020}]{Moss_2020}
Moss A., \emph{Accelerated Bayesian inference using deep learning}, 2020, \mn@doi [MNRAS]
{10.1093/mnras/staa1469}, \href
{https://doi.org/10.1093/mnras/staa1469}{496, 328}

\bibitem[\protect\citeauthoryear{{Münchmeyer} \& {Smith}}{2019}]{Munchmeyer_2019}
Münchmeyer M. and Smith K. M., 2019, \emph{Fast Wiener filtering of CMB maps with Neural Networks}
\href{https://arxiv.org/abs/1905.05846}{arXiv:1905.05846}

\bibitem[\protect\citeauthoryear{{Olvera}, {Gómez-Vargas} \& {Vázquez}}{2021}]{Olvera_2021}
Olvera J. de D. R., Gómez-Vargas I. and Vázquez J. A., 2021, \emph{Observational cosmology with Artificial Neural Networks}
\href{https://arxiv.org/abs/2112.12645}{arxiv.org/abs/2112.12645}

\bibitem[\protect\citeauthoryear{{O\'Shea} \& {Nash}}{2015}]{Oshea_2015}
{O\'Shea}, K. and {Nash}, R., \emph{An Introduction to Convolutional Neural Networks}, 2015,
\href{https://arxiv.org/abs/1511.08458}{arxiv.org/abs/1511.08458}

\bibitem[\protect\citeauthoryear{{Pal} et al.}{2023}]{Pal_2023}
Pal S., Chanda P. and Saha R., \emph{Estimation of the Full-sky Power Spectrum between Intermediate and Large Angular Scales from Partial-sky CMB Anisotropies Using an Artificial Neural Network}, 2023, \mn@doi [ApJ]
{10.3847/1538-4357/acb4ee}, \href
{https://dx.doi.org/10.3847/1538-4357/acb4ee}{945, 77}

\bibitem[\protect\citeauthoryear{{Penzias} \& {Wilson}}{1965}]{Penzias_1965}
{Penzias}, A.~A. and {Wilson}, R.~W., \emph{A Measurement of Excess Antenna Temperature at 4080 Mc/s}, 1965, \mn@doi [ApJ]
{10.1086/148307}, \href
{https://ui.adsabs.harvard.edu/abs/1965ApJ...142..419P}{142, 419}

\bibitem[\protect\citeauthoryear{{Petroff} et al.}{2020}]{Petroff_2020}
Petroff M. A., Addison G. E., Bennett C. L., Weiland J. L., \emph{Full-sky Cosmic Microwave Background Foreground Cleaning Using Machine Learning}, 2020, \mn@doi [ApJ]
{10.3847/1538-4357/abb9a7}, \href
{https://doi.org/10.3847/1538-4357/abb9a7}{903, 104}

\bibitem[\protect\citeauthoryear{{Ramanah} et al.}{2018}]{Ramanah_2018}
Ramanah, D., K., Lavaux, G. and Wandelt, B., D, \emph{Optimal and fast E/B separation with a dual messenger field}, 2018, \mn@doi [MNRAS]
{10.1093/mnras/sty341}, \href
{https://doi.org/10.1093/mnras/sty341}{476, 2825}

\bibitem[\protect\citeauthoryear{{Ramanah} et al.}{2019}]{Ramanah_2019}
Ramanah, D., K., Lavaux, G. and Wandelt, B., D, \emph{Wiener filtering and pure E/B decomposition of CMB maps with anisotropic correlated noise}, 2019, \mn@doi [MNRAS]
{10.1093/mnras/stz2608}, \href
{https://doi.org/10.1093/mnras/stz2608}{490, 947}

\bibitem[\protect\citeauthoryear{{Shallue} \& {Eisenstein}}{2023}]{Shallue_2023}
Shallue C. J. and Eisenstein D. J., \emph{Reconstructing cosmological initial conditions from late-time structure with convolutional neural networks}, 2023, \mn@doi [MNRAS]
{10.1093/mnras/stad528}, \href
{https://doi.org/10.1093/mnras/stad528}{520, 6256}

\bibitem[\protect\citeauthoryear{{Sievers} et al.}{2013}]{Sievers_2013}
Sievers J. L. et al., \emph{The Atacama Cosmology Telescope: cosmological parameters from three seasons of data}, 2013, \mn@doi [JCAP]
{10.1088/1475-7516/2013/10/060}, \href
{https://doi.org/10.1088/1475-7516/2013/10/060}{2013, 060}

\bibitem[\protect\citeauthoryear{{Smith} \& {Zaldarriga}}{2007}]{Smith_2007}
Smith, K., M. and Zaldarriaga, M., \emph{General solution to the E-B mixing problem}, 2007, \mn@doi [Phys. Rev. D]
{10.1103/PhysRevD.76.043001}, \href
{https://link.aps.org/doi/10.1103/PhysRevD.76.043001}{76, 043001}

\bibitem[\protect\citeauthoryear{{Stacey} et al.}{2018}]{Stacey_2018}
Stacey J. G. et al., \emph{CCAT-Prime: science with an ultra-widefield submillimeter observatory on Cerro Chajnantor}, 2018, \mn@doi [International Society for Optics and Photonics]
{10.1117/12.2314031}, \href
{https://doi.org/10.1117/12.2314031}{10700, 482}

\bibitem[\protect\citeauthoryear{{Tristram} et al.}{2021}]{Tristram_2021}
{Tristram, M.} et al., \emph{Planck constraints on the tensor-to-scalar ratio}, 2021, \mn@doi [A\&A]
{10.1051/0004-6361/202039585}, \href
{https://doi.org/10.1051/0004-6361/202039585}{647, A128}

\bibitem[\protect\citeauthoryear{{Wang} et al.}{2020}]{Wang_2020}
Wang G. J., Ma X. J., Li S. Y. and Xia J. Q., \emph{Reconstructing Functions and Estimating Parameters with Artificial Neural Networks: A Test with a Hubble Parameter and SNe Ia}, 2020, \mn@doi [ApJS]
{10.3847/1538-4365/ab620b}, \href
{https://doi.org/10.3847/1538-4365/ab620b}{246, 13}

\bibitem[\protect\citeauthoryear{{Zaldarriaga} \& {Seljak}}{1997}]{Zaldarriaga_1997}
Zaldarriaga, M. and Seljak, U., \emph{All-sky analysis of polarization in the microwave background}, 1997, \mn@doi [Phys. Rev. D]
{10.1103/PhysRevD.55.1830}, \href
{https://link.aps.org/doi/10.1103/PhysRevD.55.1830}{55, 1830}

\bibitem[\protect\citeauthoryear{{Zhao} \& {Baskaran}}{2010}]{Zhao_2010}
Zhao, W. and Baskaran, D., \emph{Separating E and B types of polarization on an incomplete sky}, 2010, \mn@doi [Phys. Rev. D]
{10.1103/PhysRevD.82.023001}, \href
{https://link.aps.org/doi/10.1103/PhysRevD.82.023001}{82, 023001}

\makeatother  
\end{thebibliography}
\end{document}